\documentclass[preprint,epsf]{aastex}
\usepackage{epsfig}
\def\lsim{\lower0.6ex\vbox{\hbox{$ \buildrel{\textstyle
<}\over{\sim}\ $}}}
\def\rsim{\lower0.6ex\vbox{\hbox{$ \buildrel{\textstyle
  >}\over{\sim}\ $}}}
\def\alwaysmath#1{{\ifmmode{#1}\else{$#1$}\fi}}
\def\he#1{\hbox{\alwaysmath{{}^{#1}}{\rm He}}}

\def\li#1{\hbox{\alwaysmath{{}^{#1}}{\rm Li}}}

\def\beq{\begin{equation}}
\def\eeq{\end{equation}}

\def\etal{{\it et al.}~}

\shortauthors{Olive \& Skillman}
\shorttitle{Non-Parametric Helium Abundances}

\begin{document}

\title{
A Realistic Determination of the Error on the Primordial Helium Abundance: 
Steps Toward Non-Parametric Nebular Helium Abundances
}

\author{Keith~A.~Olive}
\affil{William I. Fine Theoretical Physics Institute, \\
University of Minnesota, Minneapolis, MN 55455, USA}
\email{olive@umn.edu}

\author{Evan D. Skillman}
\affil{Astronomy Department, University of Minnesota,
      Minneapolis, MN 55455}
\email{skillman@astro.umn.edu}

\begin{abstract}
\vskip -4.5in
\begin{flushright}UMN-TH-2311/04\\FTPI-MINN-04/22\\
astro-ph/0405588 \\June 2004\end{flushright}
\vskip 3.5in

Using the WMAP determination of the baryon density, the standard model
of big bang nucleosynthesis yields relatively precise predictions
of the primordial light element abundances.  Currently there are two
significantly different observational determinations of the primordial
helium abundance, and, if only statistical errors in \he4 abundance
determinations are considered, the discrepancies between the
observational determinations and the value favored by the WMAP results
are significant.  Here we examine in detail some likely sources of
systematic uncertainties which may resolve the differences between the
two determinations. 
We conclude that the observational determination
of the primordial helium abundance is completely limited by systematic
errors and that these systematic errors have not been fully accounted
for in any published observational determination of the primordial
helium abundance. 
In principle, the observed metal-poor HII region spectra should be 
analyzed in a non-parametric way,
such that the HII region physical conditions and the helium
abundance are derived solely from the relative flux ratios of the 
helium and hydrogen emission lines.
In practice, there are very few
HII region spectra with the quality that allow this, so that most
analyses depend on assumed ranges or relationships between physical
parameters, resulting in parametric solutions with underestimated
error bars.
A representative result of our analysis yields $Y_p = 0.249 \pm 0.009$. 
However,
given that most of the spectra analyzed to date do not significantly
constrain the primordial helium abundance, we argue in favor a range
of allowed values of 0.232 $\le$ Y$_p$ $\le$ 0.258. 
This easily allows for concordance between
measurements of the baryon-to-photon ratio ($\eta$) from WMAP,
deuterium abundances, and helium abundance (although the discrepancy
with lithium remains). 

\end{abstract}


\newpage
\baselineskip=18pt
\noindent

\section{Introduction}

The standard model for cosmology can be described by a small number of
parameters, and relatively accurate determinations of many of those
parameters are now available.
Recently, cosmic microwave background experiments, most notably WMAP,  
have
determined the primordial spectrum of density fluctuations down to small
angular scales with excellent agreement with galaxy and cluster surveys
(Bennett \etal 2003; Spergel \etal 2003).
The overall curvature of the Universe is determined to be
$\Omega =
\rho/\rho_c = 1$ within a few percent, where $\rho_c =
3H^2/8\pi G_N$.
The evolution of the Universe (past, present, and future) also depends 
on
the composition of matter in the Universe: radiation, baryons, cold dark
mater, a cosmological constant or dark energy. Agreement between CMB 
measurements,
Supernovae Type I data (Riess \etal 1999; Perlmutter \etal 1999;
Tonry \etal 2003), and clustering data (Percival 2002; Tegmark 
\etal 2003) are concordant
with $\Omega_\Lambda:\Omega_m$  approximately 2:1.
Here, we are interested in the redundancies
in the determination of the baryon density,
$\Omega_B$, using both the CMB and big bang nucleosynthesis (BBN) (Walker
\etal 1991; Sarkar 1996; Olive, Steigman, \& Walker 2000; Fields \&
Sarkar 2002).

The baryon density (or the baryon-to-photon ratio,
$\eta \equiv \eta_{10}/10^{10}$) is the sole
parameter in the standard model of BBN. Prior to the recent high 
accuracy
measurements of the microwave background power spectrum, the best 
available
method for determining the baryon density of the Universe was the 
concordance
of the BBN predictions and the observations of the light element 
abundances
of D, \he3, \he4, and \li7. A high-confidence upper limit to the baryon
density has long been available (Reeves \etal 1976) from observations of
local D/H abundance determinations (giving roughly $\eta_{10} < 9.0$),
but a reliable lower bound to $\eta$, much less a precise value, has
been problematic.
Observations of each of the light elements D, \he4, and \li7 can be
used individually to determine the value of $\eta$.
Confidence in any such determination, however, relies on the concordance
of the three light isotopes.
Likelihood analyses (Fields \& Olive 1996; Fields \etal 1996; Hata \etal
1997; Fiorentini \etal 1998; Esposito \etal 2000; Cyburt, Fields, \&
Olive 2001; Burles, Nollett, \& Turner 2001) using the combined
\he4, \li7 and D/H  observations enable one to determine a 95 \% CL 
range of $5.1 < \eta_{10} < 6.7$ with a most likely value of 
$\eta_{10}  = 5.7$
corresponding to $\Omega_B h^2 = 0.021$.
However, one concern regarding the likelihood method is the
relatively poor agreement between \he4 and \li7 on the one hand and
D on the other.  The former two taken alone indicate that the most
likely  value for $\eta_{10}$ is 2.4, while D/H alone implies a best
value of  6.1.
This discrepancy may point to new physics, but could well be due to
underestimated systematic errors in the observations.
Recently, more weight has been given to the D/H determinations because of their
excellent agreement with the CMB experiments.

   In the past few years, balloon and ground-based observations have
made the first observations at multipoles $\ell \ga 200$,
where the sensitivity to $\eta$ lies, and determinations of $\eta$
at the 20 \% level became possible (Netterfield \etal 2002; Abroe \etal
2002; Pryke \etal 2002; Sievers \etal 2002; Rubi\~no-Martin \etal
2003; Goldstein \etal 2002; Benoit \etal 2002). With WMAP (Bennett \etal
2003; Spergel
\etal 2003), the CMB-based inference of the baryon-to-photon ratio is
$\Omega_B h^2 = 0.0224 \pm 0.0009$, or
$
\eta_{\rm 10,CMB} = 6.14 \pm 0.25
$
-- a precision of 4\%.
This estimate is the best-fit WMAP value,
which is sensitive mostly to WMAP alone (primarily the first
and second acoustic peaks) but does include
CBI (Sievers \etal 2002)  and ACBAR (Goldstein \etal 2002)  data on
smaller angular scales.  Note, however, that there is a small and
marginally significant difference between the values of $\eta$
derived by using only the WMAP data or using the WMAP data in
combination with other observations (Spergel et al.\ 2003) as shown
here in Figure 1.

\begin{figure}
\plotone{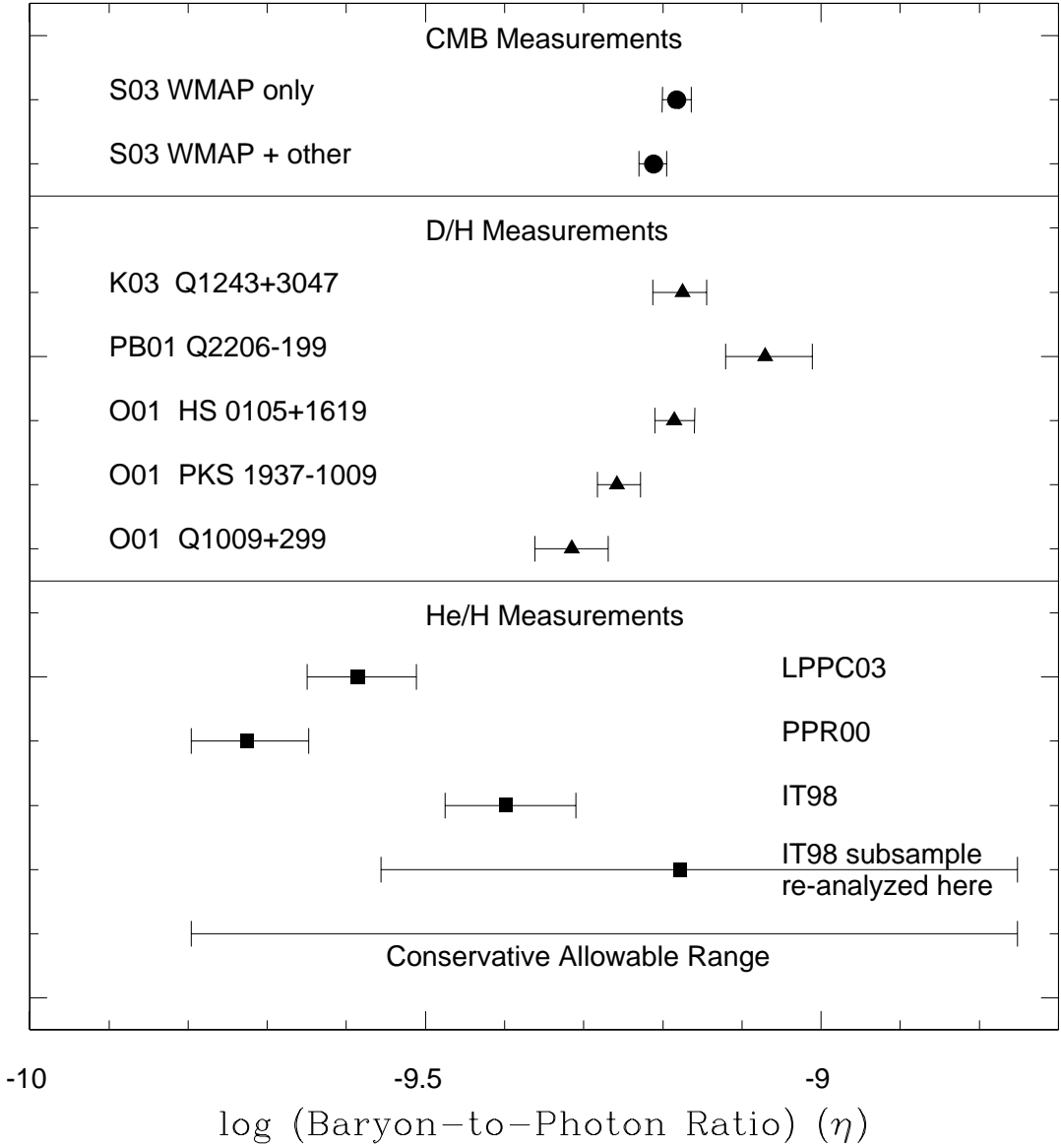}
\figcaption{
The various observational limits on the value of the baryon-to-photon 
ratio ($\eta$) individually plotted.
The value derived from He/H measurements labeled IT98 represents 
the results from 45 independent measurements while the value labeled
IT98 re-analyzed represents the value determined from a reanalysis of
the seven highest quality  measurements from that sample.
The recommended conservative range extends the range from the IT98
re-analysis to the lowest values determined by Peimbert and collaborators. 
}
\label{fig1}
\end{figure}

One can use $\eta_{\rm 10,CMB}$ as
an input to BBN calculations and predict the primordial abundances of
the light elements. This yields relatively precise abundances (Cyburt
\etal 2003):
\begin{eqnarray}({\rm D/H})_p &  = & 2.75^{+0.24}_{-0.19} \times 
10^{-5}
\nonumber \\
{\rm \he3/H} & = & 9.28^{+0.55}_{-0.54} \times 10^{-6} \nonumber \\
Y_p & = & 0.2484^{+0.0004}_{-0.0005}\nonumber \\
{\rm \li7/H} & = & 3.82^{+0.73}_{-0.60} \times 10^{-10}\nonumber \\
\end{eqnarray}
Because there are no measurements of
\he3 at very low  metallicity (i.e., significantly below solar) at this
time, a higher burden is placed on the remaining three elements.
The D/H prediction is in excellent
agreement with the average of the five best determined quasar absorption
system abundances (Burles \& Tytler 1998a,b; O'Meara \etal
2001; Pettini \& Bowen 2001; Kirkman \etal 2003) which give ${\rm D/H} =
(2.78 \pm 0.29) \times 10^{-5}$ (see Figure 1). 
In contrast, the \li7 prediction is rather high.
The results of Ryan, Norris, \& Beers (1999), give \li7/H = $1.23^{+0.34}_{-0.16}
\times 10^{-10}$ which is a factor of 3 below the WMAP value, and
almost a factor of 2 below even when systematics are stretched to
maximize the \li7 abundance (Ryan \etal 2000).
Even a recent study (Bonifacio \etal 2002)
with temperatures based on H$\alpha$ lines (considered to give
systematically high temperatures) yields \li7/H = $(2.19\pm 0.28) 
\times
10^{-10}$. Recent attempts (Coc \etal 2004, Cyburt, Fields, \& Olive 2004)
to ascribe this discrepancy to systematic uncertainties in the
relevant nuclear rates for \li7 show that this is unlikely. Of course, some of the 
discrepancy may be due to Li depletion in the stellar atmosphere (Vauclair \& Charbonnel 1998,
Pinsonneault \etal 1999, 2002).
Finally, the predicted \he4 abundance is also rather high when compared
to the large body of data on \he4 in extragalactic HII regions.

Given the concordance of the WMAP determinations with the average of
the D/H determinations, is the value of $\eta$ secure, and thus,
should the topic be considered closed?
There are a number of reasons that we feel that it is important to
resolve the differences indicated in Figure 1.
First, it is important to remember that individual cosmological
parameters cannot be derived independently from microwave background
measurements; the microwave background measurements constrain
combinations of cosmological parameters.  Thus, independent constraints
on individual parameters (or, in this case, the combination of just
two parameters - $\Omega_B h^2$) continue to be very important.
Second, the concordance of the primordial abundances of the light
elements has long been regarded as a triumph of modern cosmology,
and thus, regaining concordance remains an important goal.
Finally, there are some aspects of the $\eta$ determinations that
bear further inspection.  One example is the previously mentioned small
difference between the value derived from WMAP alone and that derived
from combining in other observations.  Another example, also shown
in Figure 1, is the fact that the error bars
on the individual D/H $\eta$ determinations, in many cases, are not
overlapping, and averaging the five best determined absorption
systems may not be the correct way to treat the data.
The dispersion in the D/H data may be a result of chemical evolution
or underlying systematic uncertainties in the observations (Fields \etal 
2001).
Thus, we feel that the problem of the concordance of the various
determinations of $\eta$ is an important one, and here we will
concentrate on the determination of the value of the primordial
helium abundance.
 
Unlike the other light element abundances, in order to be a useful
cosmological constraint, \he4 needs to be measured with a
precision at the few percent level.
Thus, the determination of the \he4 abundance has become
limited by systematic uncertainties.
To date, the most useful \he4 abundance determinations are made by
observing helium emission lines in HII regions of metal-poor dwarf
galaxies.

Izotov \& Thuan (1998, hereafter IT98) assembled a sample of
45 low metallicity HII regions, observed and analyzed in a uniform
manner, and derived a value of $Y_p$ $=$ 0.244 $\pm$ 0.002 and
0.245 $\pm$ 0.001 (with regressions against O/H and N/H 
respectively)\footnote{
This data set was recently extended to include 82 HII regions obtaining 
similar results
(Izotov \& Thuan 2004).}.
This value is significantly higher than the value of $Y_p$ $=$ 0.228
$\pm$ 0.005 derived by Pagel \etal (1992, hereafter PSTE) using nearly
identical analysis techniques.
Peimbert, Peimbert, \& Ruiz (2000, hereafter PPR00) have derived a very
accurate helium abundance for the HII region NGC 346 in the Small
Magellanic Cloud, and from this they infer a value of
$Y_p$ $=$ 0.2345 $\pm$ 0.0026.  PPR00 take a fundamentally different
approach to determining the electron temperature (compared to PSTE
and IT98).  Specifically, they use
the He~I emission lines to solve for the electron temperature
(resulting in a lower temperature than indicated by the [O~III]
lines, in line with predictions from photoionization models) and
include the effects of estimating the amplitude of temperature
fluctuations. Thus, these two different results depend not only on
different observations, but also on differences in the analyses of
the observations.  Recently, Luridiana et al.\ (2003, hereafter LPPC)
have analyzed spectra of five metal poor HII regions and, after
considering the effects of additional physical processes (e.g.,
collisional excitation of the Balmer lines), have produced a
higher (compared to PPR00) determination of Y$_p = 0.239 \pm 0.002$.  As 
shown in
Figure 1, this higher value is still significantly
lower than the value derived by IT98 or from the WMAP observations.
We are left with the conundrum in that the studies that appear to
have the higher quality observations and the more complete
physical analysis (PPR00 and LPPC) arrive at a answer which is
further from concordance with WMAP results (relative to IT98).

One approach to the problem would be to
assume that the CMB measurements are providing us with the
correct value of $\Omega_B$, and then BBN allows us to calculate the
corresponding value of Y$_p$ (Cyburt, Fields, \& Olive 2002, 2003). 
Assuming that this is correct, we may be
better able to understand the systematic effects in the determination
of Y$_p$ from nebular spectroscopy.

In an earlier paper, (Olive \& Skillman 2001, hereafter OS) we
critically examined the ``self-consistent'' approach of determining the
\he4 abundance used by IT98, and concluded that
uncertainties were systematically underestimated, especially
with regard to the treatment of underlying stellar absorption
(see also Skillman, Terlevich, \& Terlevich 1998).
This effect alone could be sufficient to explain the discrepancy
between the IT98 result and the WMAP result, but it is unlikely to
resolve the difference between the IT98 and PPR00 studies. 
Additionally, Benjamin, Skillman, \& Smits (2002) have presented new
calculations of HeI radiative transfer effects and have shown that
some of the fitting formulae introduced by IT98 had significant errors.

In this paper, we will attempt to better quantify the true
uncertainties in the individual helium determinations in
extragalactic HII regions and hopefully produce a more secure
estimate of Y$_p$.  We present an update of our minimization
procedure, adding in the ability to solve self-consistently
for the electron temperature.  The new calculations of radiative
transfer effects are also incorporated into this new code.
The goal of this paper is to explore different analysis methodologies
and to promote particular analysis techniques in order to carefully
assess the true systematic uncertainty in \he4 abundance determinations.

\section{A New Minimization Code}

In our previous work (OS), in order to make calculations directly
comparable to those of IT98, we used the H~I
emissivities calculated by Hummer \& Storey (1987), the He~I
emissivities calculated by Smits (1996), the collisional
rates of Sawey \& Berrington (1993), the resulting collisional 
corrections
calculated by Kingdon \& Ferland (1995), and the He~I radiative
transfer model of IT98 based on Robbins (1968).
Here, these are replaced and combined by new calculations from
Benjamin, Skillman, \& Smits (1999; 2002).  While the emissivity
and collisional corrections presented by Benjamin, Skillman, \& Smits 
(1999)
did not result in large changes relative to previous values, the
new radiative transfer model presented in Benjamin, Skillman, \& Smits 
(2002)
is substantially different from that of IT98.
As detailed in our previous paper, we
also allow for the possibility of underlying stellar absorption.
The details used in computing the helium abundance are given in the
Appendix.

The self consistent method we employ makes use of six He emission lines:
$\lambda$3389, $\lambda$4026, $\lambda$4471, $\lambda$5876,
$\lambda$6678, $\lambda$7026.  The ratio of each He~I emission line
intensity to H$\beta$ is compared to the theoretical ratio and
corrected for the effects of collisional excitation, fluorescence, and
underlying He~I absorption. From these line ratios,
we need to determine three physical parameters, the density,
$n$, the optical depth, $\tau$, and the equivalent width for underlying
helium absorption, $a_{HeI}$.  In OS, we had assumed an input
temperature (typically derived from [O~III] emission lines).
As pointed out in PPR00, the temperature dependences of the He~I 
emissivities
are exponential and can be both positive and negative, so that the 
relative
He~I lines can provide a strong temperature diagnostic (which will be 
more
appropriate to the He$^{+}$ zone). Thus, our new procedure
allows us to solve for the temperature as we do for the other physical
parameters.

As described in detail in OS, we use the derived He abundances
$y^+(\lambda)$ to compute the average helium abundance,
$\bar y$,
\beq
\bar y = \sum_{\lambda} {y^+(\lambda) \over \sigma(\lambda)^2} /
\sum_{\lambda}
{1 \over \sigma(\lambda)^2}
\eeq
This is a weighted average, where the uncertainty
$\sigma(\lambda)$ is found by propagating the uncertainties in the
observational  quantities stemming from the observed line fluxes (which
already contains  the uncertainty due to C(H$\beta$)), and the 
equivalent
widths.

{}From ${\bar y}$, we can define a $\chi^2$ as the deviation of the
individual He abundances $y^+(\lambda)$ from the average,
\beq
\chi^2 = \sum_{\lambda} {(y^+(\lambda) - {\bar y})^2 \over
\sigma(\lambda)^2}
\label{chi4}
\eeq
We then minimize $\chi^2$, to determine $n, a_{HeI}$, and $\tau$ and the
temperature, $T$.  Uncertainties in the output parameters are determined
by varying the outputs until $\Delta \chi^2 = 1$.

The real challenge for the new code is determining the correct 
combination of temperature and density.  While it is true, in principle,
that the He~I lines can provide a strong temperature diagnostic, in 
practice, this requires very precise emission line fluxes to pin
down the temperature to a narrow range.  The difficulty comes from 
the fact that the three strong lines which form the backbone of the
solution ($\lambda$5876, $\lambda$4471, and $\lambda$6678) all have the
same sense of temperature and density dependence, i.e., 
\beq
F_{HeI}  =  A T^{B - C n_e} 
\label{tndep}
\eeq
where A, B, and C are positive constants.  Because the three main 
lines (and $\lambda$4026) all behave in similar ways, there is a 
natural tendency towards solutions combining either high temperatures
and low densities or low temperatures and high densities.  Thus, if
temperatures are systematically overestimated (as is probably true 
if one assumes the temperature derived from the [O~III] lines), then
the solutions are forced systematically to low values of density.
The fact that two of the lines ($\lambda$3889 and $\lambda$7065)
have negative exponents in the temperature term does not provide 
as much leverage on the temperature because these two lines are 
also strongly affected by optical depth and collisional excitation 
effects. 
 
Finally, for a more robust determination of the uncertainties in the 
physical parameters and ultimately in ${\bar y}$, we perform a 
Monte-Carlo simulation of the data.
Starting with the observational inputs and their stated
uncertainties, we have generated a data set which is Gaussian
distributed  for the 6 observed He
emission lines. From each distribution, we
randomly select  a set of input values and run the
$\chi^2$ minimization. The selection  of data is repeated 1000 times. We
thus obtain a distribution of  solutions for $n, a_{HeI}$, $\tau$, and
$T$ and we compare the mean and dispersion  of these distributions with the
initial solution for these quantities.

  In simple photoionization models, HII regions divide into two 
different
ionization zones where the cooling is dominated by either O$^{+}$ or
O$^{++}$.  Typical abundance analyses usually take this into account
and derive different electron temperatures for the two different zones
(and use the different electron temperatures for deriving ionic
abundances, see Garnett 1992).  In our new code we have the 
option of using two different temperatures in determining the He/H
abundances (as done in PPR00 and LPPC, but not in IT98).  We have 
experimented with this option, and, for this study have decided
to concentrate on single temperature models.  Given the number of 
physical parameters that one is solving for and the limited number
of emission lines measured with high precision, satisfactory 
results were only achieved in cases where a relationship was
assumed between the temperatures in the low and high ionization
zones (as is usually done when calculating oxygen abundances).
Since we are advocating non-parametric solutions, we felt it 
best to stick to the simpler case of single temperature solutions. 

\section{Revisiting NGC~346 from PPR00}

    The study of NGC~346 by PPR00 emphasized the advantages (relative to
studying more distant blue compact galaxies) of (1)
no underlying absorption correction for the helium lines, (2) the
opportunity to observe multiple lines of sight, and (3) due to
a lower electron temperature, the effects of collisional excitation
on the permitted lines are small.  They also identified the disadvantage
of a larger correction for chemical evolution for the study of the
primordial helium abundance.  This study introduced the
technique of deriving the electron temperature directly from the helium 
lines.
Given the high quality of observations presented there, it is a
good starting point to compare our own analysis techniques.  Here we 
will
concentrate on their region ``A'', which provides the highest quality
spectrum.  The relevant He~I emission lines, equivalent widths, and 
errors,
taken from PPR00, are presented in Table 1.  Here we have taken the 
reddening corrected He~I
emission line fluxes and their associated errors directly from PPR00.
In analyzing these data, our main goal is to see if our analysis is 
consistent with that of PPR00.

In OS we made specific suggestions about treating the errors
in the reddening correction, and we thought it would be interesting
to revisit that question here. 
Thus, we carry out parallel analyses on a second set of data.  
The second set of data listed in Table 1 (labeled ``re-analyzed'') have 
been
derived from the original flux ratio observations of PPR00 and then 
de-reddened
using the prescription for solving for reddening and underlying H 
absorption
as described in OS.  For this second set, we have adopted the average
extinction law of Cardelli, Clayton, \& Mathis (1989; hereafter CCM). 
This is the most
commonly used astrophysical extinction law, but it differs slightly
from that of the normal extinction law of Whitford (1958) used by PPR00.
Since the extinction law is normally assumed without any associated 
errors,
we were interested in the effect of simply using a different
(although generally acceptable) extinction law.
Figure 2 shows the differences between the CCM extinction law and that
used by PPR00.  We also assign
errors of 10\% to the EW measurements (as some value of error is needed
for our minimization technique, and none is reported in PPR00).  The error
in the EW is a conservative one, but the final results do not change
significantly if the errors are varied between 10\% and 2\%.

\begin{figure}
\plotone{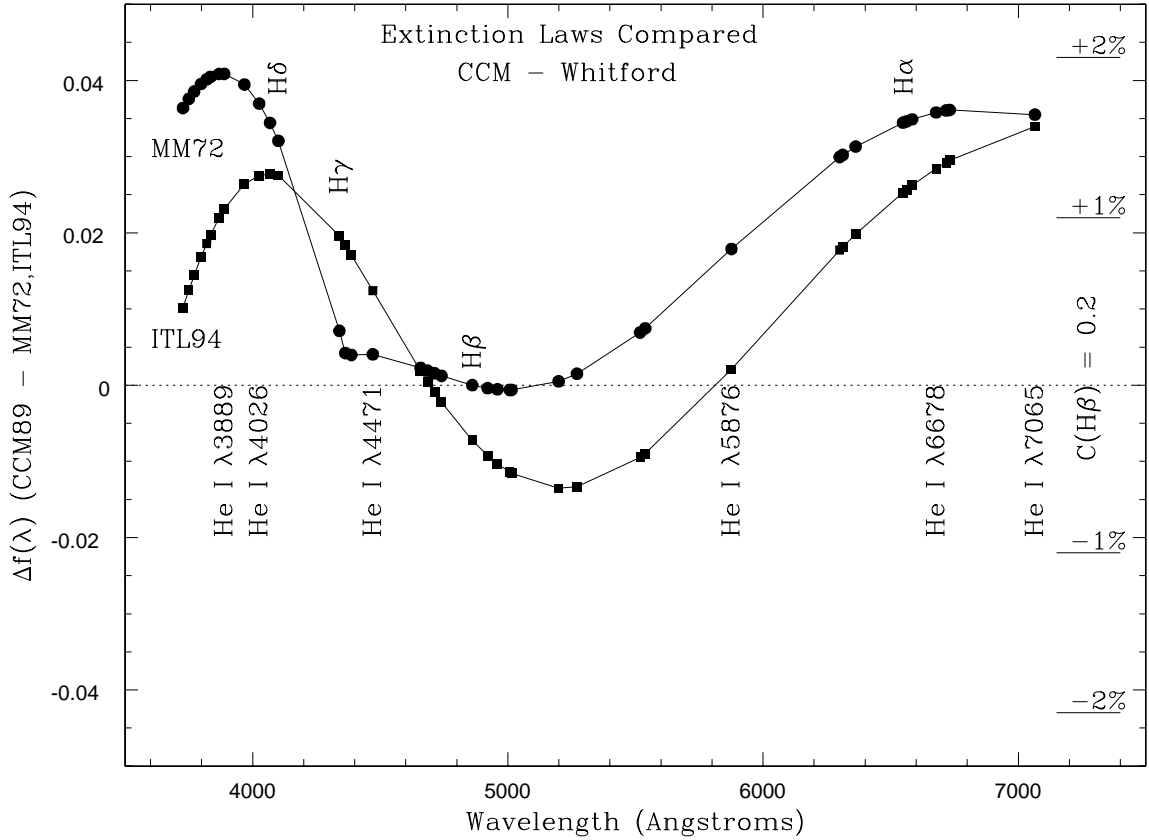}
\figcaption{
Comparisons between the extinction law of Whitford (1958) (as 
parameterized by Miller \& Mathews 1972 - MM72 - and Izotov, 
Thuan, \& Lipovetsky 1994 - ITL94) and the extinction law of 
Cardelli, Clayton, \& Mathis (1989).  The curves show
points for major HII emission emission lines.  The horizontal lines 
in the right of the figure show the resultant differences in 
dereddened emission line strengths for a logarithmic reddening at 
H$\beta$ of 0.2, which is typical of the extragalactic HII regions 
observed for helium abundance studies.
}
\label{fig2}
\end{figure}

   In what follows we refer to a simple minimization based on the
input data as a ``direct'' analysis (labeled ``D'' in the tables),
and an analysis over 1000 representations which are consistent with the
errors in the input data as a ``Monte Carlo (``MC'') analysis.''
In our first analysis, we assume a temperature of 11,920 K, the
average He~I temperature found by PPR00.  Holding this temperature fixed,
and assuming no underlying absorption,
we derive values of He$^+$/H$^+$ $=$ 0.0795 $\pm$ 0.0005  and n$_e$ 
$=$
164$^{+56}_{-47}$ which compare quite well with 0.0793 $\pm$ 0.0006
and 146 $\pm$ 50 reported by PPR00 (see Table 2).
Dropping the assumption of zero underlying absorption gives a very 
similar
result in the direct analysis, supporting PPR00's claim that there is no
underlying absorption (see Table 2).
Using a Monte Carlo analysis, we derive
He$^+$/H$^+$ $=$ 0.0801 $\pm$ 0.0011  and n$_e$ $=$ 159 $\pm$ 69, 
where
we see that the central values are only slightly changed, but the
error bars on He$^+$/H$^+$ increase by about a factor of two
(see Table 2).
The close similarity of the two results
indicates that our minimization techniques produce similar results
with high quality data and similar assumptions.  Although the PPR00
could not use the radiative transfer results of Benjamin
Skillman, \& Smits (2002), the spectrum for NGC~346 indicates very
low values of $\tau$(3889), so good agreement is expected.

\begin{figure}
\plotone{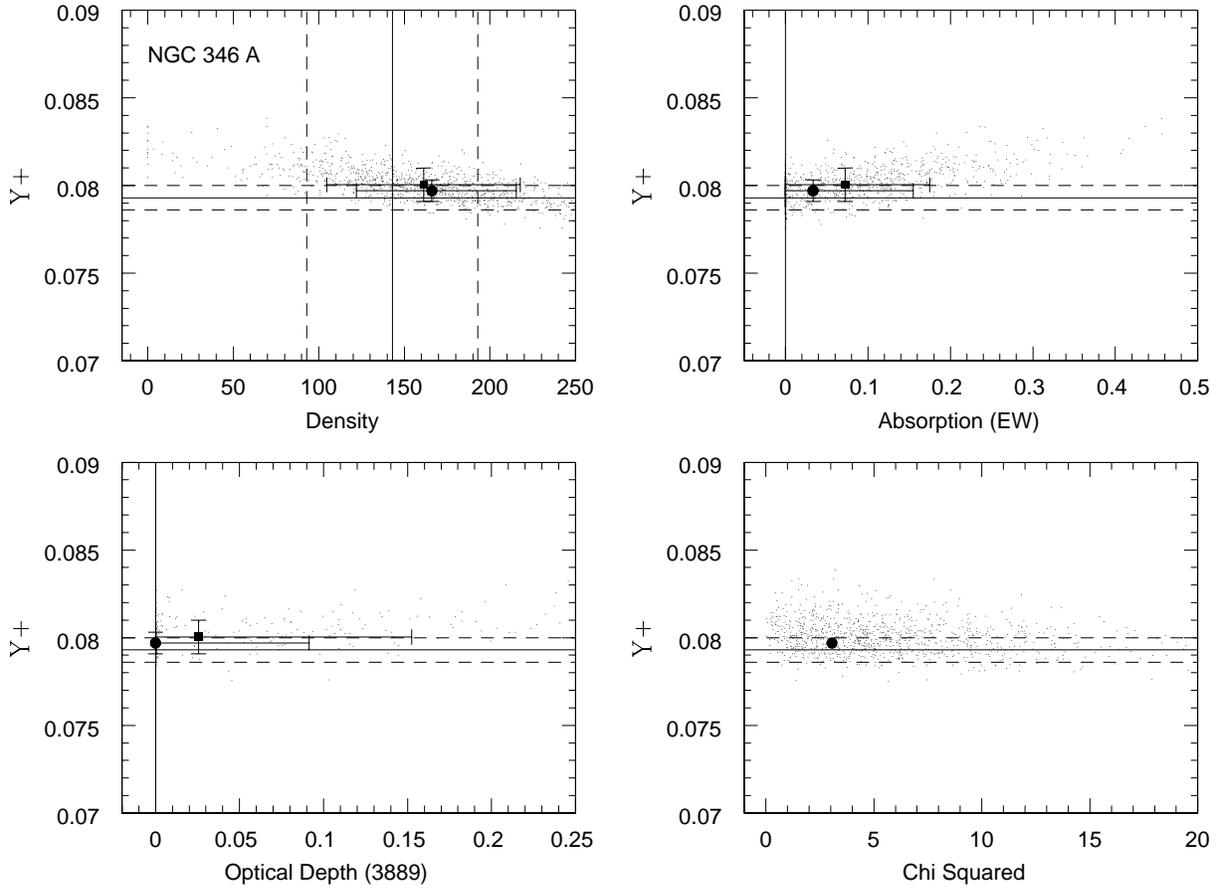}
\figcaption{
Results of modeling of 6 He~I line observations of NGC 346 from Peimbert
et al.\ (2000; PPROO).
The solid lines show the values derived by PPR00 and the dashed lines
show the  1 $\sigma$ errors on those values.
The solid circles (with error bars) show the results of our $\chi ^2$
minimization solution (with calculated  errors).
The small points show the results of Monte Carlo realizations
of the original input spectrum.
The solid squares (with error bars) show the means and dispersions
of the output values for the $\chi ^2$ minimization solutions of
the Monte Carlo realizations.
}
\label{fig3}
\end{figure}

In Figure \ref{fig3}, we show the results of our analysis for the case 
where
we have fixed the temperature and allow for  positive values of the 
underlying absorption
coefficient, $a_{HeI}$.
For this model we have used a two zone temperature model and assumed
that the temperatures in those two zones are identical to the values
derived by PPR00.  Thus, the main difference between these calculations
and those of PPR00 are dropping the assumptions of zero underlying
absorption and zero optical depth in the triplet helium lines.
Note the very good agreement between the our results (shown by the 
position
of the solid circle) and those of PPR00 (shown by the intersection of 
the
vertical and horizontal lines). The small dots show the possible 
variations in
solutions due to the Monte Carlo generation of data. The solid square 
shows
the position of the mean value of the 1000 sets of generated data.
While the mean value is consistent with both our direct solution and 
PPR00,
the uncertainty in $y^+$ is nearly doubled. 

When we allow the code to solve for the electron temperature, we
get a slightly higher value of 12,510 K
(see Table 2).  This temperature lies between
the values of T(O~III) (13,070 K) and T(O~II) (11,810 K), which would
normally be expected (Peimbert, Peimbert, \& Luridiana 2002).  
This new solution drops the $\chi^2$ value from
3.0  to 2.4, which an {\it ftest} shows is significant only at the 45\% 
confidence
level.  The solution with the higher temperature also favors a lower
density and a small amount of underlying absorption (although consistent
with zero) with a resulting He$^+$/H$^+$ ratio which is 2\% higher
than the PPR00 value.  Note that, although 2\% seems like a small
difference, the value obtained from the direct analysis allowing for
underlying absorption and deriving the electron temperature from the
helium line ratios is 3$\sigma$ higher than the PPR00 value.
The MC value is slightly higher still, but this could be
mainly due to the fact that the code requires the underlying absorption
to be positive, so that, when the value is close to zero, there is
a bias in the mean value to be larger.  The results of the MC analysis 
are shown in Figure
4.  In addition to the value of $y^+$ being larger,  the uncertainty in 
$y^+$ is
now over a factor of 2 larger than quoted in PPR00.

\begin{figure}
\plotone{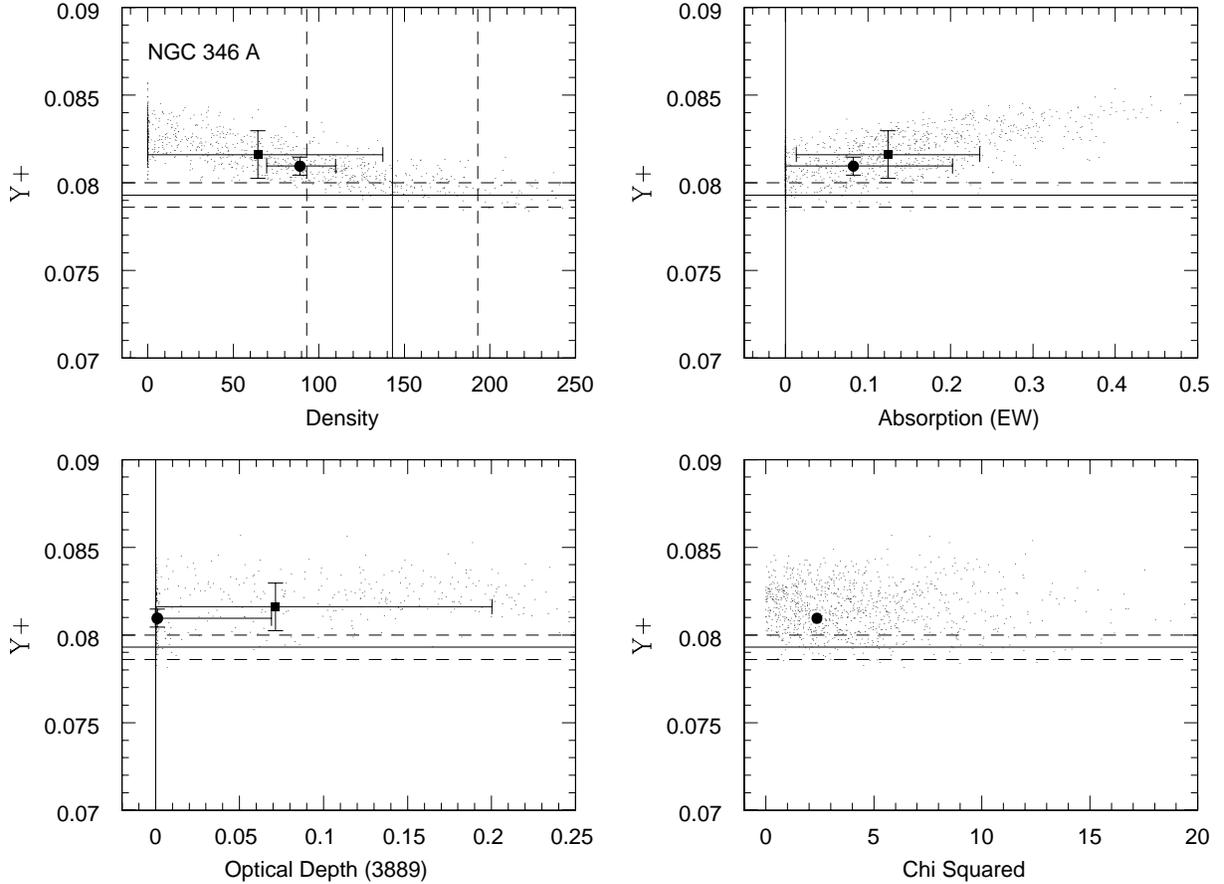}
\figcaption{
As in Fig.\ \protect\ref{fig3}, but we have assumed a single
temperature
and solved for that temperature.  Here, larger values of absorption and
lower values of density result in higher values of He/H.
The solid lines show the values derived by PPR00 and the dashed lines
show the  1 $\sigma$ errors on those values.
The solid circles (with error bars) show the results of our $\chi ^2$
minimization solution (with calculated  errors).
The small points show the results of Monte Carlo realizations
of the original input spectrum.
The solid squares (with error bars) show the means and dispersions
of the output values for the $\chi ^2$ minimization solutions of
the Monte Carlo realizations.
}
\label{fig4}
\end{figure}

As a check on the assumption of positive underlying He absorption, we 
then
ran the direct and Monte Carlo codes allowing both positive and negative
values of underlying He absorption.  The result of the direct analysis
is almost identical to that obtained with the absorption held positive
(see Table 2).
In the table, `solved +' refers to the restriction that only positive values 
of the parameter were permitted, `solved free' indicates that both positive
and negative values are allowed, and `fixed' means that the parameter
was not solved for, but rather an input value was used and held fixed.
Thus, the detection of a small amount of underlying He absorption is
not due to a bias from the code only allowing positive values.  The
higher values of the He abundance (roughly 2\% higher for the direct
method and 3\% higher for the Monte Carlo method) are probably due
to a small, but real, amount of underlying He absorption. 

Finally we ran the direct and Monte Carlo codes allowing both positive
and negative values of all three physical parameters: density, 
underlying
absorption, and optical depth while still solving for electron 
temperature (last lines in Table 2). 
Note that negative values of density and
optical depth are non-physical. 
Now the favored value for the underlying absorption is
almost very close to zero.
However, it is very interesting that
the free solution has favored negative (non-physical) values of the
optical depth.
In this solution,
because of the negative value for the optical depth, the density
has jumped up to 427 (mainly because of the inverse dependences on
density and optical depth for the $\lambda$7065 line), resulting in
a He$^+$/H$^+$ value which is even lower than the PPR00 value. 
The favored negative values of the optical depth indicate that either
our prescription for the treatment of radiative transfer on the 
helium line strengths is not appropriate for NGC~346 or that the value
for $\lambda$3889 (the main driver of the optical depth solution) is
somewhat in error (or both).

{}From this exercise we conclude that the error bars on the derivation
of individual HII region parameters are underestimated for several
reasons.  First, as concluded in OS, unless a Monte Carlo analysis of
the data is performed, the resulting errors of a single minimization
exercise are likely to be underestimates. 
In the case for which we solve for temperature
and absorption, our MC uncertainty is over a factor of 2 larger
than that quoted by PPR00.  Second, by making different
reasonable assumptions, different central values of the helium abundance
are obtained (and these values differ by several standard deviations if
the errors are calculated in the direct manner).
For NGC~346A these differences are only a few percent because the 
quality
of the data constrains the results to a relatively narrow range. However,
typical HII region spectra used for deriving helium abundances are not
even close
to the quality of the NGC~346A spectrum from PPR00.  Third, by allowing 
the
code to examine non-physical values of certain parameters there is an
indication that perhaps the prescription for treating the radiative
transfer is not completely appropriate.  This implies a systematic
uncertainty which has not been accounted for in the errors analysis.

  In Table 2 we also report the results of re-analyzing the NGC~346A
spectrum using our favored prescriptions for the solution for reddening
and underlying hydrogen absorption and the appropriate error
propagation (we label these entries ``re-an'').  Our analysis 
finds a similar value for C(H$\beta$)
 $=$ 0.174 $\pm$ 0.008  (compared to 0.15 $\pm$ 0.01 for PPR00), and
no evidence of underlying hydrogen
absorption (formally $-$0.6 $\pm$ 0.5 \AA ). 
As seen in Table 1, there are small differences between the PPR00
relative line fluxes and those resulting from our re-analysis, which
are mostly due to the differences in the adopted reddening curve
(and some small differences
might be expected simply due to round-off errors).  As expected,
only small differences are found when we analyze these helium
ratios (as shown in Table 2).  There is, however, one interesting difference.  
When we solve for all of the parameters including temperature,
the temperature is higher (although not as high as the [O~III] temperature)
with the result that
the density falls and the helium abundance rises by 2\% relative
to that found in the similar analysis with the PPR00 data.
Both the original data and the re-analyzed data favor negative values
of the optical depth.
It appears that even with excellent quality
spectra, one is never far from unsatisfactory results, and that
one needs exquisite quality data to insure that the best minimization
values favor physically meaningful values.

A detailed inspection of Table 2 indicates that the values derived
for the physical parameters and the helium abundance depend on the
assumptions that are made in the analysis.  While the variations are
of order 2 -- 3\% for most of the helium abundance determinations,
this is very significant for determination of $\eta$, and of order
the discrepancies found in the literature.
For example, the PPR00 value of $y^+ = 0.0793$ corresponds to a
\he4 mass fraction of 0.2403, from which they inferred a primordial
abundance of 0.2345 which in turn corresponds to $\eta_{10} = 1.9$.
Our value (using the re-analyzed data) of $y^+ = 0.0828$ corresponds 
to a
\he4 mass fraction of 0.2483. Using the same slope (dY/dZ) yields 
a primordial
abundance of 0.2425 which in turn corresponds to $\eta_{10} = 3.5$.
Thus, one sees a difference of nearly a factor of two in the
derived value of $\eta_{10}$ simply from using a different set
of assumptions.
 
Figure 5 underscores this point.
Here we have plotted the He$^+$/H$^+$ ratios and errors
derived for each line as a result of our direct method which solves
simultaneously for helium abundance, temperature, density, underlying
absorption and optical depth under the assumption that all are positive.
The solid and dotted lines across the bottom indicate the solution from
PPR00.  The significant difference is due primarily to assumptions made in
analyzing the observations.  In other words, an independent analysis,
using reasonable assumptions, produces a value of  He$^+$/H$^+$ $=$
0.0828 $\pm$ 0.0008, which is over 4\% (or 6$\sigma$) higher than
the PPR00 value.  We are not claiming here that this solution is
superior to that produced by PPR00, only that it is a completely
satisfactory solution but different.  These systematic differences
need to be accounted for in estimating errors in the final helium
abundance.

\begin{figure}
\plotone{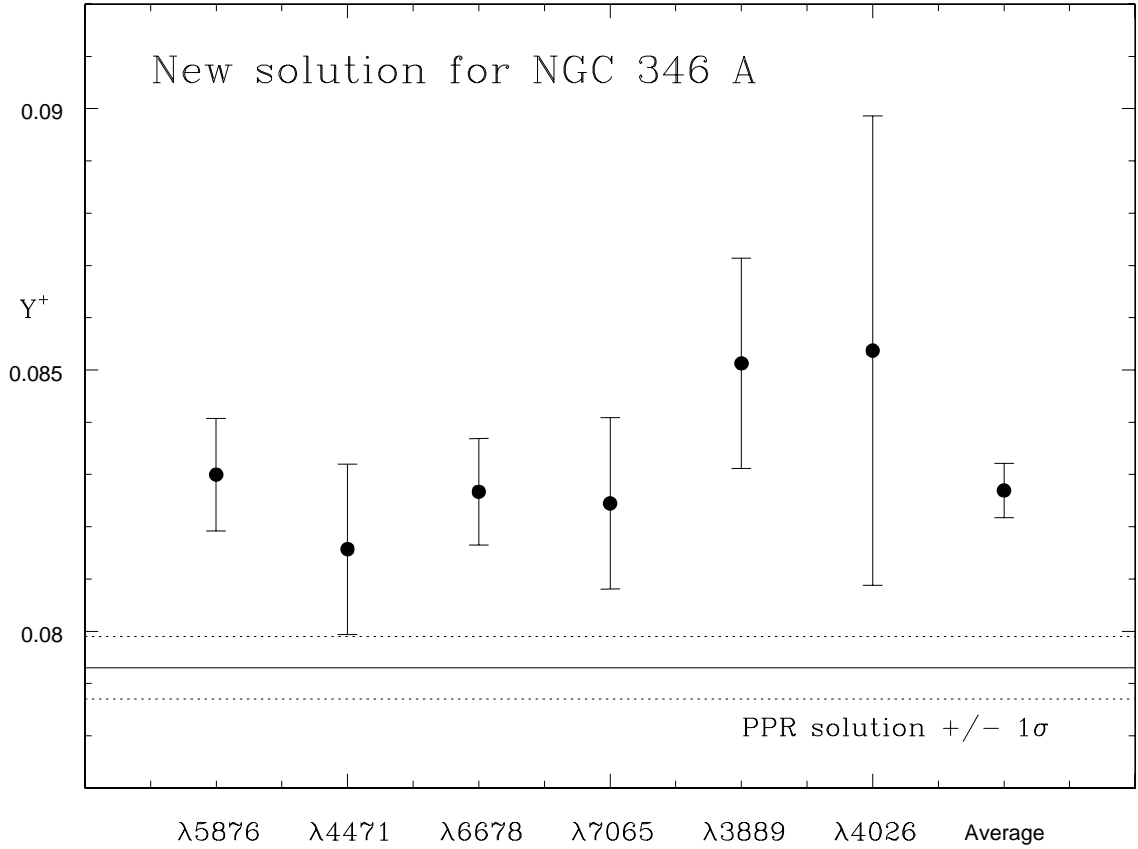}
\figcaption{
Results of modeling of 6 He~I line observations of NGC 346 A from Peimbert
et al.\ (2000; PPROO).  Here we show the He$^+$/H$^+$ ratios and errors
derived for each line as a result of our direct method which solves
simultaneously for helium abundance, temperature, density, underlying
absorption and optical depth under the assumption that all are positive.
The solid and dotted lines across the bottom indicate the solution from
PPR00.  The significant difference is due primarily to assumptions made in
analyzing the observations.
}
\label{fig5}
\end{figure}

What is the ``best'' value for He$^+$/H$^+$ in NGC~346A?  We favor
an estimation which is as close to a non-parametric solution as 
possible.
Thus, we feel that solving for electron temperature from the helium
lines and allowing only for non-negative values of underlying absorption 
is
preferred\footnote{If the underlying absorption is negative, then our 
assumption
that it is roughly equal at all wavelengths is no longer true.  That is 
one of the reasons
we also search for solutions where $a_{HeI}$ (as well as the other 
parameters)
is allowed to be negative.}.
This is not a truly non-parametric solution as we are still dependent on
assumptions concerning the relative importance of underlying helium 
absorption
and optical depth on the different helium lines.  For our analysis of
the PPR00 data for NGC 346A, this means that we favor 0.0828 $\pm$ 0.0008,
which, compared with the result of PPR00 (0.0793 $\pm$ 0.0006) appears
extremely high, and would even appear to rule out their value.  At this
point we are not ready to interpret the results in that way.  Our main
conclusion is that the systematic error of analysis technique is
clearly dominant, and that this error, when reported, has been
grossly underestimated in the past.

\section{Revisiting Select Targets from IT98}

    In is now useful to revisit selected targets from IT98 to
better assess the robustness of their results.  Many of the targets
in IT98 are not of the quality necessary for high accuracy helium
abundance determinations.  For example, the majority are not of 
sufficient quality to detect the $\lambda$4026 line which provides 
a strong constraint on the underlying He~I absorption.  Thus, we have
chosen to re-analyze only a select few targets which meet three
criteria.  First we select only the IT98 targets with
EW(H$\beta$) $\ge$ 200\AA .  This cut is made in order to minimize the 
correction due to underlying absorption as well as to improve the 
signal-to-noise ratio in faint lines such as $\lambda$4026 (the PPR00 
value of EW(H$\beta$) for NGC 346A is 250).  The value of 200 \AA\ is
identified by Izotov \& Thuan (2004; IT04) as a desirable cut-off.  
This selection screens out
35 of the 45 IT98 spectra.  
Next we select targets with oxygen abundances less than 20\% of the solar
value (4.6 $\times$ 10$^{-4}$; Asplund \etal\ 2004).
This cut minimizes the uncertainty induced by our lack of knowledge of 
the functional form of the relationship between \he4 and O/H (c.f., 
Olive, Steigman, \& Walker 1991; Pilyugin 1993; Carigi et al.\ 1995; 
Bradamante, Matteucci, \& D'Ercole 1998;
Fields \& Olive 1998). This selection screens out 17 of the 45 IT98 
spectra.  Finally we select out targets with radial velocities inside 
of the ranges 728 $\pm$ 100 km s$^{-1}$ and 1032 $\pm$ 100 km s$^{-1}$ 
(to avoid contamination of He~I $\lambda$5876 by Na~I absorption; cf.\
Dinerstein \& Shields 1986). This selection screens out 10 of the 45
IT98 spectra.   Clearly several of the targets fail more than one of
the screens.   In the end, we find there are seven targets which 
pass all three screens, and hereafter we will refer
to this subsample as the ``high quality'' sample.  

    We proceed in the same manner as our analysis of NGC 346A, i.e.,
we analyze the reported reddening corrected He~I line strengths
directly and also re-analyze the data starting from the raw 
flux ratio measurements.
    For our analysis, we require the EW of the He~I lines.  In
most cases these are not reported, so we have estimated the strength
of the continuum (from published spectra) and used the EW(H$\beta$)
to derive these EWs.
This is, of course, less than desirable, and, at first, we
hesitated to carry this out.  However, since the primary effect
is on the value of the underlying absorption, and since it is the
ratio of the lines strengths that matters here, it turns out that
the answers have a relatively low sensitivity to these estimated
values.  We assigned uncertainties to the estimated EW of 10\%.
In fact, changes of these values of up to 20\%, in most cases,
have little effect on the derived helium abundances.  In cases
where we have re-analyzed the data by deriving our own reddening
and underlying absorption corrections, we have also estimated the
EW of the Balmer lines in the same way.

\subsection{SBS 0335-052}

A spectrum for SBS 0335-052 was reported in IT98 (and therefore
we do not need to make any adjustments to be consistent with
the IT98 analysis as will be the case for other targets).
When we re-analyze the IT98 spectrum for SBS 0335-052
using our favored prescriptions for the
solution for reddening and underlying hydrogen absorption,
we obtain a value of C(H$\beta$) $=$ 0.121 $\pm$ 0.014,
in good agreement with IT98 value of 0.13, and we find a higher
value for the underlying hydrogen absorption of 1.3 $\pm$ 0.9 \AA\ 
(to be compared with IT98 value of 0.4 \AA ).
The resulting helium line strengths are reported in Table 3.
There are only small differences between the IT98 values and the
values resulting from our re-analysis.  The one noticeable difference
is for $\lambda$3889, where our value is 9\% higher due primarily
to the correction for underlying H absorption.

Table 4 shows the the results of the full set of analyses of the IT98 spectrum
for SBS 0335-052. Our constrained analyses with the assumption of no
underlying He absorption find much higher values of the optical depth in
the in helium lines ($\approx$ 5 vs.\ 1.7) and this results in lower
densities but almost no difference in the
value of the He abundance.  However, when we allow for
underlying absorption, there is significant evidence for this and this
results in higher helium abundances.  
The $\chi^2$ drops significantly with the allowance for underlying
absorption and the helium abundance rises significantly.  Given
the large variation in helium abundance values for the different
degrees of constraint, it would appear that this spectrum of SBS0335-052 
is not
suitable for constraining the primordial helium abundance.

By solving for the temperature, significantly lower temperatures 
(with significantly larger uncertainties)
and higher densities are found, with the resultant much lower He
abundances.  The derived temperatures are so much lower than
the [O~III] temperatures such that these solutions are most likely
not to be believed.  The very high values of $\tau$(3889) are robust,
and it is unlikely that our prescription for correcting for optical
depth effects is valid in this regime.

Although SBS 0335-052 is an ideal target from the viewpoint of its
exceptionally low O/H abundance, the very high value of optical depth 
means that this is probably a less than suitable object (at least 
in the highest surface brightness knot) for determining helium
abundance.  This is reflected in the fact that the error in He$^+$/H$^+$
in SBS 0335-052 is the largest (in both absolute terms and relative
terms) of the seven objects from IT98 that we have re-analyzed.

\subsection{NGC 2363A}

The spectrum for NGC 2363A analyzed in IT98 was first reported in 
Izotov, Thuan, \& Lipovetsky (1997; ITL97).
In ITL97, an electron temperature of 15,100 $\pm$ 100 K, an oxygen 
abundance of 0.78 $\pm$ 0.01 $\times$ 10$^{-4}$, and a helium abundance
of Y $=$ 0.245 $\pm$ 0.005 were reported.  IT98 re-analyzed this spectrum,
and, primarily because of a difference in the estimated electron 
temperature, the oxygen abundance changed to 0.71 $\pm$ 0.01 $\times$ 
10$^{-4}$ (a 7 $\sigma$ decrease), and the  
helium abundance changed to Y $=$ 0.2456 $\pm$ 0.0008 (an 84\% decrease
in the uncertainty).  That such different values can be reported from
the same spectrum by the same researchers gives one an impression
of the size of the systematic errors.  For our comparison analysis, we will
assume an [O~III] electron temperature of 15,800 $\pm$ 100 K, which would 
correspond to the scale used in IT98. 

When we re-analyze the IT98 spectrum for NGC~2363A 
using our favored prescriptions for the
solution for reddening and underlying hydrogen absorption,
we obtain a value of C(H$\beta$) $=$ 0.117 $\pm$ 0.003,
in good agreement with the ITL97 value of 0.11, and we find a slightly 
lower value for the underlying hydrogen absorption of 1.4 $\pm$ 0.3 \AA\ 
(to be compared with ITL97 value of 1.8 \AA ).
The resulting helium line strengths are reported in Table 3.
There are only small differences between the IT98 values and the
values resulting from our re-analysis (all consistent with round-off
errors). 

Table 5 shows the results of our analyses of the HeI line ratios.
In all analyses where their presence is allowed, we find evidence
for significant amounts of underlying absorption and optical depth
effects (which were both assumed to be absent in the IT98 analysis).
Holding the electron temperature fixed at the [O~III] temperature and
assuming no underlying absorption reveals values of $\tau$(3889)
$\approx$ 2.  When underlying absorption is allowed for, the solutions 
favor significant amounts (the $\chi ^2$ drops by a factor of 6).  Finally,
solving for the electron temperature yields a lower temperature.  Note
that the value of the derived helium abundance varies by 8\% 
depending on the set of assumptions used in the derivation.  Perhaps
the most remarkable thing about this re-analysis is that our value 
derived by allowing for underlying absorption and optical depth effects
and solving for the temperature produces a value nearly identical to
the IT98 value although our uncertainty (from the MC analysis) is 4.6 times
larger.

\subsection{SBS 0940+544N}

The spectrum for SBS 0940+544N analyzed in IT98 was reported in ITL97.
In ITL97, an electron temperature of 19,000 $\pm$ 300 K, an oxygen
abundance of 0.30 $\pm$ 0.01 $\times$ 10$^{-4}$, and a helium abundance
of Y $=$ 0.246 $\pm$ 0.008 were reported.  IT98 re-analyzed this spectrum,
and, primarily because of a difference in the estimated electron
temperature, the oxygen abundance changed to 0.27 $\pm$ 0.01 $\times$
10$^{-4}$, and the helium abundance changed to Y $=$ 0.2500 $\pm$ 0.0047. 
For our comparison analysis, we will
assume an [O~III] electron temperature of 20,200 $\pm$ 300 K, which would
correspond to the scale used in IT98.

When we re-analyze the IT98 spectrum for SBS 0940+544N
using our favored prescriptions for the
solution for reddening and underlying hydrogen absorption,
we obtain a value of C(H$\beta$) $=$ 0.048 $\pm$ 0.023,
in good agreement with the ITL97 value of 0.05, but, formally, we find 
a negative value (though consistent with 0) 
for the underlying hydrogen absorption of $-$1.4 $\pm$ 1.4 
\AA\  to be compared with ITL97 value of 0.6 \AA.  This presents
a problem concerning how to proceed with the analysis.  Since the 
solution is formally consistent with no underlying absorption, we
assume a value of 0 $\pm$ 1.4 \AA\  in our analyses.
The resulting helium line strengths are reported in Table 3.
There are only small differences between the IT98 values and the
values resulting from our re-analysis (all consistent with round-off
errors).

Table 6 shows the results of our analyses of the HeI line ratios.
Our analyses are generally in agreement with IT98 in that low values of
underlying absorption, optical depth, and density are found.
However, perhaps the most interesting part of our re-analysis can be
seen when non-physical values are allowed for.  Specifically, negative 
values of underlying absorption and optical depth are favored.  These
non-physical values indicate that there must be some problem with either
the spectrum or our simple assumptions used when analyzing it.

\subsection{MRK 193}

The spectrum for MRK 193 analyzed in IT98 was first reported in 
ITL94.
In ITL94, an electron temperature of 15,500, an oxygen
abundance of 0.70 $\pm$ 0.02 $\times$ 10$^{-4}$, and a helium abundance
of Y $=$ 0.253 $\pm$ 0.007 were reported.  IT98 re-analyzed this spectrum,
and, primarily because of a difference in the estimated electron
temperature, the oxygen abundance changed to 0.64 $\pm$ 0.01 $\times$
10$^{-4}$, and the helium abundance changed to Y $=$ 0.2478 $\pm$ 0.0037.
For our comparison analysis, we will
assume an [O~III] electron temperature of 16,200 $\pm$ 200 K, which would
correspond to the scale used in IT98.

When we re-analyze the IT98 spectrum for MRK 193
using our favored prescriptions for the
solution for reddening and underlying hydrogen absorption,
we obtain a value of C(H$\beta$) $=$ 0.292 $\pm$ 0.020,
in good agreement with the ITL94 value of 0.260, but, formally, we find
a negative value for the underlying hydrogen absorption of $-$1.7 $\pm$ 
1.0 \AA\ (to be compared with ITL94 value of 1.9 \AA ).  As in the case
of SBS 0940+544N, this again presents
a problem concerning how to proceed with the analysis.  In this case, the 
solution is not formally consistent with zero underlying absorption, but we
are forced to assume a value of 0 $\pm$ 1.0 \AA\  in our analyses.
The resulting helium line strengths are reported in Table 3.
There are only small differences between the IT98 values and the
values resulting from our re-analysis, except for the difference in
I($\lambda$3889).

Table 7 shows the results of our analyses of the HeI line ratios.
Our analyses are generally in agreement with IT98 in that low values are
found for underlying absorption and relatively high densities of density.  
However, non-zero values of optical
depth are favored, and strongly so in the re-analyzed data.
When the temperature is solved for, lower values are favored.
Overall, once again the impression is that the assumptions of the
analysis strongly affect the derived value of Y.

\subsection{SBS 1159+545}

The spectrum for SBS 1159+545 analyzed in IT98 was first reported in ITL94.
IT98 re-analyzed this spectrum upon discovering errors made in the 
data reduction reported in ITL94.
Since the full spectrum and analysis for SBS 1159+545 was reported in IT98 
we do not need to make any adjustments for consistency.
When we re-analyze the IT98 spectrum for
SBS 1159+545  using our favored prescriptions for the
solution for reddening and underlying hydrogen absorption,
we obtain a value of C(H$\beta$) $=$ 0.065 $\pm$ 0.016,
in good agreement with the IT98 value of 0.06, and we find a
value for the underlying hydrogen absorption of 0.2 $\pm$ 1.1 \AA\ 
(smaller than but consistent with the IT98 value of 0.6 \AA ). 
The resulting helium line strengths are reported in Table 3, and,
in general, there is very good agreement.

Table 8 shows the results of several analyses of the IT98 spectrum for
SBS 1159+545.  When we fix the temperature to the IT98 value and assume
no underlying He absorption, we find a value of the helium abundance
consistent with that of IT98 and also an optical depth of zero,
consistent with IT98.  Interestingly, when we re-analyze the
spectrum, the helium line strengths change only slightly, but even
in the constrained analysis, the favored density has fallen, the
optical depth is non-zero, and the favored helium abundance has risen
by 2\% (while the $\chi^2$ is almost identical).

In the constrained solutions, there is never any evidence of underlying
He absorption in the SBS 1159+545 spectrum.  The less constrained 
solutions
show why.  There, negative values of underlying absorption are strongly 
favored. 
Indeed, in the unconstrained solutions, negative values for
both underlying He absorption and optical depth are favored, resulting
in very high values of density, relatively low values of temperature, 
and thus, unrealistically low values of the He$^+$/H$^+$ ratio.
Although the SBS 1159+545 spectrum is fairly high signal/noise,
it would appear that a robust measurement of the helium abundance
may be suspect for this spectrum. 

\subsection{Haro 29 }

The spectrum for Haro 29 analyzed in IT98 was reported in ITL97.
In ITL97, an electron temperature of 15,400 $\pm$ 100 K, an oxygen
abundance of 0.65 $\pm$ 0.01 $\times$ 10$^{-4}$, and a helium abundance
of Y $=$ 0.246 $\pm$ 0.005 were reported.  IT98 re-analyzed this spectrum,
and, primarily because of a difference in the estimated electron
temperature, the oxygen abundance changed to 0.59 $\pm$ 0.01 $\times$
10$^{-4}$, and the helium abundance changed to Y $=$ 0.2509 $\pm$ 0.0012.
For our comparison analysis, we will
assume an [O~III] electron temperature of 16,180 $\pm$ 100 K, which would
correspond to the scale used in IT98.

When we re-analyze the IT98 spectrum for
Haro 29  using our favored prescriptions for the
solution for reddening and underlying hydrogen absorption,
we obtain a value of C(H$\beta$) $=$ 0.001 $\pm$ 0.004,
in good agreement with ITL97 value of 0.0, and we find a
value for the underlying hydrogen absorption of 2.7 $\pm$ 0.3 \AA\ 
(to be compared with IT98 value of 2.5 \AA ).
The resulting helium line strengths are reported in Table 3,
and, in general, there is very good agreement.

Table 9 shows the the results of several analyses of the IT98 spectrum
for Haro 29.  Our constrained analyses with the assumption of no
underlying He absorption find evidence for non-zero optical depth in
the in Helium lines and this results in higher densities and thus,
lower values of the He abundance.  However, when we allow for
underlying absorption, there is very significant evidence 
(the $\chi^2$ drops roughly a factor of 6 with the allowance 
for underlying absorption) and this
results in higher helium abundances.  
When we solve for the temperature, the value is very close to
that obtained from the [O~III] lines, which is reassuring, and the
solutions do not change significantly when we allow for positive or
negative values for the underlying absorption.  However, 
for our least constrained solutions, we find the
very discouraging result that very different solutions are found
with higher values of the optical depth, higher values of the
underlying He absorption, negative densities, higher electron
temperatures and very high helium abundances.  The slightly
lower $\chi^2$ for these solutions are probably only a warning
of how loosely constrained even the best solutions are.

  LPPC03 have analyzed the IT98 spectrum of Haro 29 and derive a
value of He$^+$/H$^+$ $=$ 0.078.  There are two large differences
between their treatment and the one presented here.  First, a
much larger value of the reddening is adopted and second,
underlying He absorption is assumed to be zero. In all of our solutions,
we find this to be a poor assumption.  

\subsection{SBS 1420+544}

A spectrum for SBS 1420+544 was reported in IT98 (and therefore
we do not need to make any adjustments to be consistent with
the IT98 analysis as for most of the targets).
When we re-analyze the IT98 spectrum for
SBS 1420+544  using our favored prescriptions for the
solution for reddening and underlying hydrogen absorption,
we obtain a value of C(H$\beta$) $=$ 0.167 $\pm$ 0.013,
in good agreement with IT98 value of 0.16, and we find a
value for the underlying hydrogen absorption of 0.4 $\pm$ 0.9 \AA\ 
(to be compared with IT98 value of 0.0 \AA ). 
The resulting helium line strengths are reported in Table 3,
and, in general, there is very good agreement with the IT98 values.

Table 10 shows the the results of several analyses of the IT98 spectrum 
for SBS 1420+544.  SBS 1420+544 is interesting in that it is one of the
few HII regions for which IT98 find definite signs of significant
optical depth.  Our constrained solutions with underlying absorption
fixed at zero find higher values for the optical depth, and this
results in larger values of the density and lower values of the
He$^+$/H$^+$ ratio.  When we allow for underlying He absorption,
there is only weak or no evidence of this.  However, when we allow
for the temperature to be solved for, significantly larger temperatures
result, giving lower densities and higher helium abundances.
Using our re-analyzed spectrum and solving for temperature,
the resulting helium abundance is significantly higher (by 4\%) than 
IT98.

Given the evidence for the large optical depth in the helium lines,
and the large uncertainty in our prescription for
correcting for this, it would seem that again, a robust measurement
of the helium abundance is difficult for this spectrum.

\subsection{Summary of Re-analysis of IT98 Spectra}

   A summary of our re-analysis of the IT98 spectra is given in
Table 11.  Several differences between the results of the IT98 
analysis and our re-analysis are immediately apparent.
First, the physical parameters T, N, $\tau$(3889), and He(ABS) are
all significantly less well constrained in our re-analysis.  In some cases, this
is because some of these values were assumed to be zero with no
associated errors, and in other cases we believe that the errors 
were underestimated.  The larger uncertainties in the physical 
parameters naturally translate into larger uncertainties in the
helium abundances.

  Table 11 shows a larger dispersion in the re-analyzed He abundances, 
with a slight bias toward higher values.  Note that in the discussions
of the individual objects we raised reservations concerning the 
suitability of several of these spectra for deriving very high 
accuracy helium abundances.  Specifically, SBS 0940+544N and MRK 193
both favored negative (non-physical) values for underlying hydrogen
absorption.  SBS 0335-052 and SBS 1420+544 both have relatively large
values of $\tau$(3889), and we doubt that the simple prescription
for correcting for optical depth effects will be accurate in this
regime.  Although we specifically selected objects with high emission
line equivalent width, two objects (NGC 2363 A and Haro 29) show 
significant evidence for underlying helium absorption, and thus, 
are dependent on our assumption of the behavior of the absorption
equivalent widths on a line-by-line basis.

  In sum, even the spectra that were selected specifically to be the 
highest quality provide only nominal constraints on the physical
conditions in the HII regions, and thus, carry significant errors
on their helium abundances compared to what is necessary to 
constrain the value of the primordial helium abundance.

\section{Revisiting Select Targets from IT04}

   Recently, Izotov \& Thuan (2004; IT04) have presented new observations
of 33 HII regions in order to provide a better metallicity baseline and
therefore stronger constraints on $dY/dZ$.  They combine these new 
observations with those of IT98 to conduct an analysis of 82 HII regions.
We have studied the new spectra presented by IT04 to determine if there
are new observations which satisfy our criteria for inclusion in a
``high quality'' data sample, and, in fact, there are seven targets
that do satisfy the criteria enumerated at the beginning of \S 4.
These seven targets, in order of increasing metallicity are: 
J~0519+0007, HS~2236+1344, HS~0122+0743, HS~0837+4717,
CGCG~007-025, HS~0134+3415, and HS~1028+3843.  We performed a 
complete analysis of these seven targets following the method used
in \S 4.

\begin{figure}
\plotone{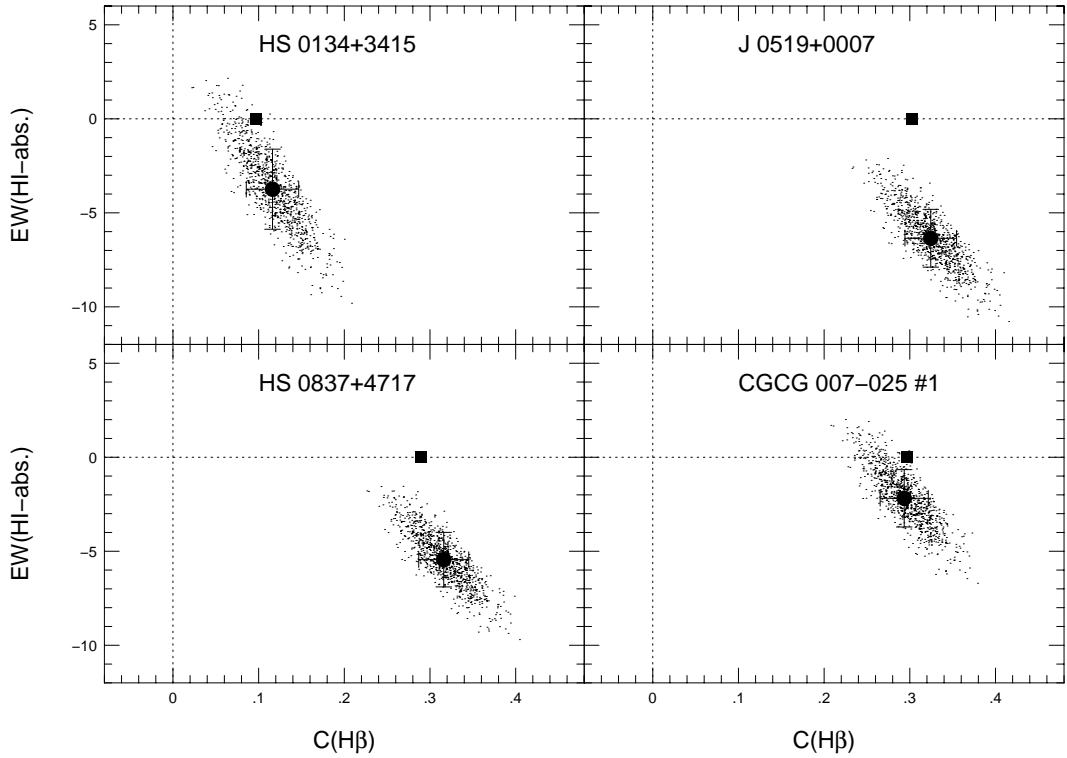}
\figcaption{
Solutions for the value of reddening, c(H$\beta$), and underlying
Balmer absorption, EW(HI-abs.), for four of
the ``high quality'' data points from IT04.
In all four cases, we find that the ``best'' solution requires
significantly negative values of EW(HI-abs.) which is probably
not physically realistic.
The small points show the results of our Monte Carlo modeling and
the large circles with error bars show our best values with errorbars.
The square points are the solutions reported by and used by IT04.
}
\label{fig6}
\end{figure}

   Unfortunately, we discovered a serious problem with many of the
observations reported in IT04, which render them questionable for use in our analyses.  
When we solve for the reddening and underlying Balmer absorption
in the spectra from IT04, we often encounter significantly negative
values of underlying Balmer absorption.  This is found in 4 of the 7
cases identified as potential ``high quality'' sample targets.
Figure 6 shows the solutions for these four sources.  The solutions
for underlying absorption are found to be negative ranging 
from 1.4 $\sigma$ for CGCG~007-025 to 4.1 $\sigma$ for J~0519+0007.
As can be seen from Figure 6, in all cases IT04 derived values of
0 \AA\ for the underlying Balmer ratios. This may be due to limiting the
solution to physically meaningful, i.e., positive values.

   As noted before, it was problematic that two of the targets in 
the IT98 ``high quality'' data sample showed negative values for 
underlying Balmer absorption (SBS~0940+544N at the 1.0 $\sigma$ level
and Mrk 193 at 1.7 $\sigma$), but the prevalence in the new sample
cannot be ignored.  In fact, roughly half of the new observations
reported in IT04 indicate zero underlying Balmer absorption, indicating
that this problem exists throughout the dataset.  
If we look at the object in the ``high quality'' data sample 
with the most negative value of underlying  Balmer absorption (J~0519+0007)
we find that the reddening
corrected values for the blue Balmer lines are all significantly too
high relative to the theoretical values (i.e., H$\gamma$/H$\beta$,
H$\delta$/H$\beta$, H9/H$\beta$, and H10/H$\beta$ are $+$4.4 $\sigma$,
$+$5.1 $\sigma$, $+$0.4 $\sigma$, and $+$1.8 $\sigma$ respectively).
In fact, the values for the blue Balmer lines before reddening 
correction are all close to their theoretical values, indicating
very low reddening to this target.  Brief inspection of the other
targets reveals a similar pattern, indicating that the problem
may be solely with the H$\alpha$ line (in the sense that the 
H$\alpha$ line is too strong) and this could point to a problem
with the sensitivity calibration in that wavelength range.
On the other hand, J~0519+0007 is at a Galactic latitude of
$-$20$\degr$ and according to Schlegel et al.\ (1998) should have
an extinction of A$_V$ $\sim$ 0.4 mag.  If there is no problem with
the sensitivity calibration, then the effect could be due to 
atmospheric differential refraction (cf., Filippenko 1982) even 
thought the object was observed at an airmass of 1.2.
An alternative explanation could be that these spectra indicate
the collisional enhancement of the H$\alpha$ (see discussion 
in Skillman \& Kennicutt 1993), but it seems unlikely that we
would see it in so many sources and not find a strong correlation
with electron temperature.

  Until this problem is solved, we have reservations concerning the 
inclusion of the spectra from IT04 in our study of the primordial helium 
abundance.  In Table 12 we have reported our re-analysis of the ``high
quality'' sample of IT04.  Unfortunately, IT04 do not report their
derived values for $n$ and $\tau$(3889) so direct comparisons are not
possible, but it appears that many of the characteristics from Table
11 are also found in Table 12.  In our analysis, several of the targets
show relatively large values of $\tau$(3889) and significant evidence
for underlying HeI absorption.

\section{Towards a Primordial \he4 Abundance}

Given the different possible assumptions one can make
regarding the treatment of the underlying absorption, optical
depth, or temperature, one can justifiably wonder whether it is indeed
possible to extract a reliable value of the primordial abundance 
of \he4 from these observations.  It is our hope that by striving
to free our solutions from as many parametric constraints as 
possible and by considering all identifiable sources of error
that we can now produce a value of the primordial helium 
abundance with reasonable error estimates.

\begin{figure}
\plotone{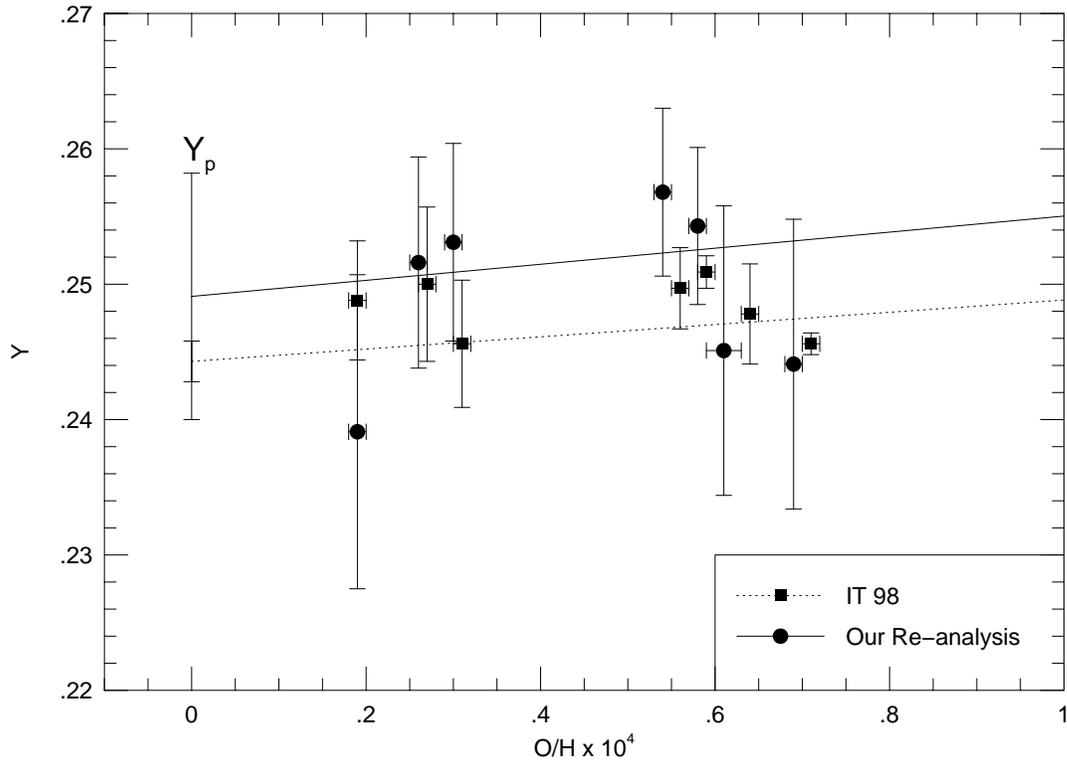}
\figcaption{
A comparison of the results for the best IT98 targets and our re-analysis
of the spectra for those targets.
The regression and the intercept with errors (Y$_p$) shown for IT98 is 
that resulting from the full sample of IT98 while the regression for
the re-analysis is based on only the seven ``high quality'' targets. 
}
\label{fig7}
\end{figure}

If we take our preferred solutions based on the re-analyzed data
(where we use the Monte Carlo technique and solve for temperature
and restrict all parameters to take positive values) 
for the seven ``high quality'' targets from IT98, then a regression
of y$^+$ on O/H yields an intercept of:
\beq
Y_p = 0.2491 \pm 0.0091
\eeq
The corresponding value for the baryon-to-photon ratio is
\beq
\eta_{10} = 6.64^{+11.1}_{-3.82}
\eeq
easily in good agreement with the CMB determination of $\eta$.
Figure 7 shows a comparison between the original IT98 values and
our re-analysis for the seven ``high quality'' targets.  It also
compares the results of the above regression with the original
value obtained by IT98 for the full set of 45 targets.
The error on Y$_p$ derived above and shown in Figure 7 is 4.5
times larger than derived by IT98.  
One could argue that we
should drop the two observations which yielded negative
Balmer absorption.  In this case, our regression for five points
would yield, $Y_p = 0.2470 \pm 0.0114$.  While the intercept
has decreased slightly, the uncertainty has increased
making the 1 $\sigma$ upper limit virtually identical.

Obviously, some of the increase
in the error is due to the decrease in the sample size from 45 
to 7, and some of the increase in the error is due to a larger error
on the slope due to a smaller metallicity baseline.  However,
a good part of the increase in the error is due to the larger
error on the individual points.  
Additionally, some of 
the decrease in the intercept error gained by 
extending the metallicity baseline to higher
metallicities is misleading because it is dependent on the
questionable assumption of a perfectly linear relationship between 
Y and O/H. 

We can investigate the effect of limiting the sample size by
including some of the IT04 data points.
If we include the three ``high quality'' targets from IT04 which do
not show evidence of negative underlying H~I absorption, using 
all eight points which show no sign of negative H~I absorption we obtain:
\beq
Y_p = 0.2502 \pm 0.0093
\eeq
The corresponding value for the baryon-to-photon ratio is
\beq
\eta_{10} = 7.48^{+12.9}_{-4.45}
\eeq
again, easily in good agreement with the CMB determination of $\eta$.
Finally, if we include all seven ``high quality'' targets from IT04 
together with all seven from IT98, we obtain:
\beq
Y_p = 0.2504 \pm 0.0073
\eeq
The corresponding value for the baryon-to-photon ratio is
\beq
\eta_{10} = 7.63^{+9.1}_{-3.95}
\eeq
also easily in good agreement with the CMB determination of $\eta$.

Note that although the error on Y$_p$ is declining with the 
increase in number of targets, it is not decreasing as the 
square root of the number of targets.  It appears that there is 
a floor in the uncertainty which is likely greater than 0.005. 

In Figure 1 we summarize our view of the present situation.  Clearly
the true uncertainty on the primordial helium abundance is much
larger than indicated in previous studies.  We cannot rule out values
as high as the those achieved with our re-analysis of the IT98 
observations.  Neither can we rule out values as low as those
derived by PPR00. 
Thus, perhaps the best we can say is that the primordial \he4 abundance
lies within the ``conservative allowable range'' indicated in Figure 1.
\beq
0.232 \le Y_p \le 0.258
\eeq
This extended range in $Y_p$ corresponds to
\beq
1.8 \le \eta_{10} \le 18
\eeq
Thus, at present, there is no conflict with the CMB inferred value of $Y_p$.  

\section{Discussion}

One of the key results of the preceding exercise is that despite having
apparently 
high quality data, the final \he4 abundance for any given HII region
will depend critically on assumptions concerning the appropriate electron
temperature, the presence of underlying absorption, and the treatment of 
the optical depth effects.  Many of the systems analyzed 
here show significant evidence for underlying absorption which leads to an 
increase in the derived \he4 abundance.  Additionally, many of the
systems show significant evidence for non-zero values of $\tau$(3889)
which relies on an untested formula for correction.

Furthermore, as one can see from the results for individual HII regions 
shown in the
tables, the spread in the derived \he4 abundances can be far greater
than the derived error for a given solution.  This is a clear sign that
systematic uncertainties are dominating the error budget. 
While some of the systematics such as optical depth and underlying 
absorption
move the \he4 abundance in opposite directions, it makes no sense
to suppose that these systematic uncertainties cancel each 
other out and therefore can be ignored as argued in IT04.  
Nor does it make sense to
continuously average literature data to beat down the statistical errors,
once again in complete neglect of the systematic uncertainties
as attempted in Dmitriev, Flambaum \& Webb (2004).
In this context, we stress that there is no evidence for 
a discrepancy between BBN predictions of \he4 using the WMAP
baryon density and observational determinations of the \he4 abundance
as frequently reported in the literature.

Inherent degeneracies between the parameters allow for very different 
solutions with equally acceptable
$\chi^2$. This was the main point of our previous work (OS). Perhaps 
most troubling is the degeneracy in the parameter space identified 
here between temperature and density.
As we noted throughout our analysis, IT98 have fixed the temperature to 
the OIII temperature which is systematically high 
(Peimbert, Peimbert, \& Luridiana 2002).  
However, as one can see from eqs.\  (A2),
changes in temperature can be compensated by changes in the density
allowing for the possibility of very different results due to various 
corrections.
This may explain the puzzling result that the dispersion in the IT98
helium abundances is actually smaller than their formal errors.
By assuming that the OIII temperatures are appropriate to the entire
He$^{+}$ emission zone, the densities have been biased 
to unrealistically low values resulting in higher values of 
the helium abundance.  On the other hand, by
assuming that underlying He absorption is negligible, the derived helium
abundances have been biased to lower values.  The narrow range of
derived helium abundances may be a result of these biases in 
the analysis. 

There are other potentially important sources for systematic uncertainty
such as temperature fluctuations (Steigman, Viegas, \& Gruenwald 1997; 
Sauer \& Jedamzik 2002),
ionization corrections (Ballantyne, Ferland, \& Martin 2000;
Viegas, Gruenwald, \& Steigman 2000;
Gruenwald, Steigman \& Viegas 2002; Sauer \& Jedamzik 2002),
and the collisional excitation of hydrogen emission lines (Skillman \& 
Kennicutt 1993; Stasi{\' n}ska \& Izotov 2001, LPPC).
However, it is our impression that these are small compared to the 
uncertainties that we have focused on here. 

    From our analyses, it appears that the high EW
regions may be more prone to high values of $\tau$(3889), which makes them
less suitable candidates.  This could be a real problem for identifying
new candidates for high precision helium abundances.  If derived helium
abundances depend on our simple model for helium optical depth effects, 
we have a systematic uncertainty which, at present, is not testable.
At this point, the prudent thing to do is to exclude all targets with
relatively high values of $\tau$(3889) from any analysis sample (as 
opposed to trying to correct for such effects).

   We suggest that the following efforts will result in future progress 
on obtaining a more accurate value for the primordial helium abundance.
On the observation side:

(1) Significantly higher quality spectra are needed for almost all
of the regions studied to date.  Concentrating on those objects which
appear to have higher EW(H$\beta$) and low $\tau$(3889), in addition
to low O/H abundance, will be important. 

(2) It is still important to
search for new targets.  Obviously the targets which satisfy the above
criteria are rare (Terlevich, Skillman, \& Terlevich 1996). 

(3) Observations along multiple lines of sight to 
individual targets will allow testing of our analysis techniques as the
physical conditions are changing while (hopefully) the helium abundances
are not (e.g., Skillman et al.\ 1994).

(4) Observations are needed at higher spectral resolution and over a
larger wavelength range.  The higher resolution (better than 1 \AA )
will allow the profiles of the underlying HeI absorption to be observed,
and thus allow for better corrections.   The larger wavelength range
will allow measurements of additional physical parameters (i.e., the
temperature in the lower ionization zone from [O~II] lines) which will
place stronger constraints on the derived physical parameters.  Long-slit
double spectrographs provide the best tool for this work.

On the theoretical side, work remains in order to better
characterize the uncertainties in the current atomic data and the 
prescriptions by which the observations are converted into 
abundances (e.g., Porter, Bauman, \& Ferland 2003). 

In sum, echoing our introduction, pinning down the primordial helium
abundance to a higher accuracy remains a worthwhile goal.  To significantly
decrease the uncertainty in the primordial helium abundance will 
require spectra of metal poor HII regions of a quality exceeding most 
of those present in the literature today and  analyses that 
take into account the several sources of systematic errors which have 
been identified so far. 

\acknowledgments
We would like to thank R.\ Benjamin, D.\ Garnett, R.\ Kennicutt, D.\ Kunth, 
H.\ Lee, V.\ Luridiana, B.\ Pagel,
A.\ Peimbert, M.\ Peimbert, G.\  Shields, J.\  Shields, G.\ Steigman,
S.\ Viegas, E.\ Terlevich, and R.\ Terlevich,  for informative and
valuable discussions. The work of KAO is supported in  part by DOE grant
DE-FG02-94ER-40823. EDS is grateful for partial support from a NASA
LTSARP grant No. NAG5-9221 and the University of Minnesota.
EDS is also grateful to the Institute of Astronomy at the
University of Cambridge where much of this work was done while on
sabbatical leave from the University of Minnesota.

\appendix

\section{Helium Line Ratios}

We again start with a set of observed quantities: line intensities
$I(\lambda)$ which include the reddening correction previously
determined in OS and its associated uncertainty and the equivalent width
$W(\lambda)$. The Helium line intensities are
scaled to $H\beta$  and the singly ionized helium abundance is given by
\beq
y^+(\lambda) = {I(\lambda) \over I(H\beta)} {F_\lambda(n_e,T) \over
f_\lambda(n_e,T,\tau)} \left({W(\lambda) + a_{HeI}
\over W(\lambda)}\right)
\label{y+}
\eeq
where $F_\lambda(n_e,T)$ is the theoretical emissivity scaled
to $H\beta$ and the optical depth function $f_\lambda$ contains the
collisional correction.  The expression (\ref{y+}), also contains a
correction  factor for underlying stellar absorption, parameterized now
by $a_{HeI}$  Thus $y^+$ implicitly
depends on four unknowns, the electron density, $n$, $a_{HeI}$,
$\tau$, and $T$.

To be definite, we list here the necessary components in expression
(\ref{y+}). The theoretical emissivities scaled
to $H\beta$ are taken from Benjamin, Skillman, \& Smits (1999):
\begin{eqnarray}
F_{3889} & = & 0.904 T^{-0.173 - 0.00054 n_e} \nonumber \\
F_{4026} & = & 4.297 T^{0.090 - 0.0000063 n_e} \nonumber \\
F_{4471} & = & 2.010 T^{0.127 - 0.00041 n_e} \nonumber \\
F_{5876} & = & 0.735 T^{0.230 - 0.00063 n_e} \nonumber \\
F_{6678} & = & 2.580 T^{0.249 - 0.00020 n_e} \nonumber \\
F_{7065} & = & 12.45 T^{-0.917} /(3.4940 -
(0.793-.0015 n_e + 0.000000696 n_e^2)T)
\end{eqnarray}
  Our expressions for the
optical depth function which includes the collisional correction
is also taken from Benjamin, Skillman, \& Smits (1999).  We list them
here
for completeness. They are:
\begin{eqnarray}
f(3889) & = & 1 + (\tau/2) \left[ -0.106 + (5.14 \times 10^{-5} - 4.20
\times 10^{-7} n_e + 1.97 \times 10^{-10} n_e^2)T) \right] \nonumber \\
f(4026) & = & 1 + (\tau/2) \left[ 0.00143 + (4.05 \times 10^{-4} + 
3.63
\times 10^{-8} n_e )T)
\right]
\nonumber\\
f(4471) & = & 1 + (\tau/2) \left[ 0.00274 + (8.81 \times 10^{-4} - 
1.21
\times 10^{-6} n_e )T)
\right] \nonumber \\
f(5876) & = & 1 + (\tau/2) \left[ 0.00470 + (2.23 \times 10^{-3} - 
2.51
\times 10^{-6} n_e )T)
\right] \nonumber \\
f(6678) & = & 1  \nonumber \\
f(7065) & = & 1 + (\tau/2) \left[ 0.359 + (- 3.46 \times 10^{-2} - 
1.84
\times 10^{-4} n_e + 3.039 \times 10^{-7} n_e^2)T)
\right]
\end{eqnarray}

Finally we note that we discovered two sign errors in OS concerning the theoretical
Balmer line ratios.  In Eq. A.3 of that paper, the last two expressions each contain an error.
The correct expressions should be: 
$X_T(4340)  =  -0.01655(\log T_4)^2 + 0.02824 \log T_4 + 0.468$
and
$X_T(4101)  =  -0.01655(\log T_4)^2 + 0.02159 \log T_4 + 0.259$.


\clearpage
\begin{deluxetable}{lcc}
\tablenum{1}
\tablewidth{0pt}
\tablecaption{He Emission Lines and EW for NGC~346 from PPR00 \label{tbl-1}}
\tablehead{ 
\colhead{He~I line} & \colhead{I($\lambda$)/I(H$\beta$)}     &
\colhead{EW (\AA)}} 
\startdata 
 & From PPR00 & \\
\hline     
$\lambda$5876 & 0.1064 $\pm$ 0.0012 & 46   $\pm$ 4.6  \\  
$\lambda$4471 & 0.0384 $\pm$ 0.0006 & 8.5  $\pm$ 0.85  \\
$\lambda$6678 & 0.0296 $\pm$ 0.0003 & 14.7 $\pm$ 1.47  \\
$\lambda$7065 & 0.0211 $\pm$ 0.0002 & 11.1 $\pm$ 1.11  \\
$\lambda$3889 & 0.0940 $\pm$ 0.0017 & 16.1 $\pm$ 1.61  \\                        $\lambda$4026 & 0.0185 $\pm$ 0.0006 & 3.16 $\pm$ 0.32  \\
\hline
 & PPR00 (re-analyzed) & \\
\hline     
$\lambda$5876 & 0.1053 $\pm$ 0.0013 & 46   $\pm$ 4.6  \\  
$\lambda$4471 & 0.0388 $\pm$ 0.0005 & 8.5  $\pm$ 0.85  \\
$\lambda$6678 & 0.0296 $\pm$ 0.0002 & 14.7 $\pm$ 1.47  \\
$\lambda$7065 & 0.0210 $\pm$ 0.0002 & 11.2 $\pm$ 1.12  \\
$\lambda$3889 & 0.0987 $\pm$ 0.0022 & 16.5 $\pm$ 1.65  \\
$\lambda$4026 & 0.0189 $\pm$ 0.0007 & 3.10 $\pm$ 0.31  \\
\enddata

\end{deluxetable}

\clearpage
\begin{deluxetable}{lcccccc}
\tabletypesize{\footnotesize}
\tablenum{2}
\tablewidth{0pt}
\tablecaption{Analysis of NGC~346 from PPR00 \label{tbl-2}}
\tablehead{
\colhead{Analysis} &
\colhead{He$^+$/H$^+$}     &
\colhead{T$_e$ $<(He II)>$}     &
\colhead{N$_e$}     &
\colhead{He~I ABS EW (\AA)}     &
\colhead{$\tau$(3889)}     &
\colhead{$\chi ^{2}$}}
\startdata
PPR00     & 0.0793 $\pm$ 0.0006 & 11,920 $\pm$ 370 & 146 $\pm$ 50  & 0  & 0 & ... \\
         & (solved)            & (solved)         & (solved $+$)  & (fixed) & (fixed) \\
\hline
D        & 0.0795 $\pm$ 0.0005 & 11,920 $\pm$ 370 & 164 $^{+56}_{-47}$ &
            0                      & 0.00 $^{+0.09}_{-0.00}$ & 3.1  \\ 
D (re-an)& 0.0797 $\pm$ 0.0007 & 11,920 $\pm$ 370 & 149 $^{+51}_{-46}$ &
            0                      & 0.00 $^{+0.06}_{-0.00}$ & 10.0  \\
     & (solved)    & (fixed) & (solved $+$)  & (fixed)  & (solved $+$) \\
\hline
D        & 0.0797 $\pm$ 0.0006 & 11,920 $\pm$ 370 & 163 $^{+54}_{-49}$ &
           0.03 $^{+0.12}_{-0.03}$ & 0.00 $^{+0.09}_{-0.00}$ & 3.0   \\  
MC       & 0.0801 $\pm$ 0.0011 & 11,920 $\pm$ 370 & 159 $\pm$ 69 &
           0.07 $^{+0.11}_{-0.07}$ & 0.03 $^{+0.13}_{-0.03}$ & ... \\
D (re-an)& 0.0797 $\pm$ 0.0007 & 11,920 $\pm$ 370 & 148 $^{+53}_{-46}$ &
           0.00 $^{+0.07}_{-0.00}$ & 0.00 $^{+0.06}_{-0.00}$ & 10.0  \\
MC (re-an)& 0.0800 $\pm$ 0.0009 & 11,920 $\pm$ 370 & 152 $\pm$ 63 &
           0.03 $^{+0.09}_{-0.03}$  & 0.00 $^{+0.01}_{-0.00}$ & ... \\
     & (solved)    & (fixed) & (solved $+$)  & (solved $+$) & (solved $+$) \\
\hline
D        & 0.0810 $\pm$ 0.0005 & 12,510 $\pm$ 190 & 89 $^{+21}_{-19}$ &
           0.08 $^{+0.12}_{-0.08}$ & 0.00 $^{+0.07}_{-0.00}$  & 2.4  \\  
MC       & 0.0816 $\pm$ 0.0014 & 12,650 $\pm$ 580 & 67 $^{+69}_{-67}$ &
           0.12 $\pm$ 0.11     & 0.06 $^{+0.14}_{-0.06}$ & ... \\
D (re-an) & 0.0827 $\pm$ 0.0005 & 13,460 $\pm$ 230 & 0.1 $^{+12}_{-0.1}$ &
           0.07 $^{+0.12}_{-0.07}$ & 0.00 $^{+0.04}_{-0.00}$  & 3.7   \\
MC (re-an)& 0.0828 $\pm$ 0.0008 & 13,420 $\pm$ 280 & 2 $^{+19}_{-2}$ &
           0.09 $\pm$ 0.09     & 0.01 $^{+0.04}_{-0.01}$ & ... \\
     & (solved)    & (solved) & (solved $+$)  & (solved $+$) & (solved $+$) \\
\hline
D        & 0.0810 $\pm$ 0.0005 & 12,510 $\pm$ 190 & 89 $^{+21}_{-20}$ &
           0.08 $^{+0.12}_{-0.11}$ & 0.00 $^{+0.07}_{-0.00}$  & 2.4 \\
MC       & 0.0820 $\pm$ 0.0014 & 12,810 $\pm$ 540 & 36 $^{+94}_{-36}$   &
           0.13   $\pm$ 0.13   & 0.13 $\pm$ 0.13  & ... \\
D (re-an)& 0.0827 $\pm$ 0.0005 & 13,460 $\pm$ 230 & 0.3  $^{+12}_{-0.2}$ &
           0.07   $\pm$ 0.12   & 0.00 $^{+0.04}_{-0.00}$ & 3.7  \\
MC (re-an)& 0.0827 $\pm$ 0.0009 & 13,430 $\pm$ 260 & 1 $^{+11}_{-1}$  &
           0.07 $^{+0.18}_{-0.07}$ & 0.01 $^{+0.06}_{-0.01}$  & ... \\
     & (solved)    & (solved) & (solved $+$)  & (solved free) & (solved $+$) \\
\hline
D        & 0.0777 $\pm$ 0.0005 & 11,530 $\pm$ 150 & 427 $^{+47}_{-42}$ &
           $-$0.05 $\pm$ 0.11    & $-$0.47 $\pm$ 0.07 & 1.9 \\  
MC       & 0.0813 $\pm$ 0.0084 & 13,000 $\pm$ 3,580 & 376 $\pm$ 415 &
          0.10 $\pm$ 0.36   & 0.07 $\pm$ 1.14 & ... \\
D (re-an)& 0.0829 $\pm$ 0.0005 & 13,820 $\pm$ 240 & 22 $\pm$ 16 &
         0.06 $\pm$ 0.12 & $-$0.24 $\pm$ 0.07 & 2.3 \\
MC (re-an)& 0.0836 $\pm$ 0.0051 & 14,080 $\pm$ 1,770 & 72 $\pm$ 223 &
         0.09 $\pm$ 0.24   & $-$0.16 $\pm$ 0.56 & ... \\
     & (solved)    & (solved) & (solved free)  & (solved free) & (solved free) \\
\enddata
\end{deluxetable}


\begin{deluxetable}{lcccccccc}
\tabletypesize{\footnotesize}
\rotate
\tablenum{3a}
\tablewidth{0pt}
\tablecaption{He Emission Lines and EW for IT98 ``High Quality'' Data Sample
\label{tbl-3a}}
\tablehead{
\colhead{He~I line} & \colhead{SBS 0335 -025} && \colhead{NGC 2363A} &&
\colhead{SBS 0940+544} && \colhead{MRK 193}  \\ 
& \colhead{I($\lambda$)/I(H$\beta$)} & \colhead{EW (\AA)}
& \colhead{I($\lambda$)/I(H$\beta$)} & \colhead{EW (\AA)}
& \colhead{I($\lambda$)/I(H$\beta$)} & \colhead{EW (\AA)}
& \colhead{I($\lambda$)/I(H$\beta$)} & \colhead{EW (\AA)}
}
\startdata
 & & & & from IT98 & \\
\hline
$\lambda$5876 & 0.098 $\pm$ 0.002 & 35.3 $\pm$ 3.5 & 0.106 $\pm$ 0.001 & 63.0 $\pm$ 6.3 & 0.099 $\pm$ 0.003 & 36.7 $\pm$ 3.67 & 0.111 $\pm$ 0.002 & 38.1 $\pm$ 3.81  \\
$\lambda$4471 & 0.037 $\pm$ 0.002 & 6.1  $\pm$ 0.6 & 0.039 $\pm$ 0.001 & 10.9 $\pm$ 1.1 & 0.037 $\pm$ 0.002 & 6.6  $\pm$ 0.66 & 0.039 $\pm$ 0.002 & 6.7  $\pm$ 0.67  \\
$\lambda$6678 & 0.026 $\pm$ 0.001 & 12.5 $\pm$ 1.3 & 0.029 $\pm$ 0.001 & 24.0 $\pm$ 2.4 & 0.026 $\pm$ 0.001 & 12.8 $\pm$ 1.28 & 0.028 $\pm$ 0.001 & 14.5 $\pm$ 1.45  \\
$\lambda$7065 & 0.039 $\pm$ 0.001 & 22.9 $\pm$ 2.3 & 0.029 $\pm$ 0.001 & 29.7 $\pm$ 3.0 & 0.026 $\pm$ 0.002 & 12.3 $\pm$ 1.23 & 0.032 $\pm$ 0.002 & 19.1 $\pm$ 1.91  \\
$\lambda$3889 & 0.064 $\pm$ 0.004 & 7.5 $\pm$ 0.8  & 0.088 $\pm$ 0.001 & 16.8 $\pm$ 1.7 & 0.109 $\pm$ 0.007 & 13.1 $\pm$ 1.31  & 0.100 $\pm$ 0.005 &  9.1 $\pm$ 0.91  \\
$\lambda$4026 & 0.016 $\pm$ 0.002 & 2.0 $\pm$ 0.2  & 0.016 $\pm$ 0.001 & 3.3  $\pm$ 0.3 & 0.021 $\pm$ 0.002 & 2.7  $\pm$ 0.27  & 0.018 $\pm$ 0.002 & 2.0 $\pm$ 0.20  \\
\hline
 & & & & (re-analyzed) & \\
\hline
$\lambda$5876 & 0.0977 $\pm$ 0.0020 & 35.3 $\pm$ 3.5 & 0.1056 $\pm$ 0.0010 & 63.0 $\pm$ 6.3  & 0.0997 $\pm$ 0.0031 & 36.7 $\pm$ 3.67  & 0.1122 $\pm$ 0.0028 & 38.1 $\pm$ 3.81  \\
$\lambda$4471 & 0.0370 $\pm$ 0.0021 & 6.1  $\pm$ 0.6 & 0.0391 $\pm$ 0.0010 & 10.9 $\pm$ 1.1  & 0.0375 $\pm$ 0.0020 & 6.6  $\pm$ 0.72  & 0.0399 $\pm$ 0.0022 & 6.7  $\pm$ 0.73  \\
$\lambda$6678 & 0.0264 $\pm$ 0.0010 & 12.5 $\pm$ 1.3 & 0.0284 $\pm$ 0.0009 & 24.0 $\pm$ 2.4  & 0.0270 $\pm$ 0.0020 & 12.8 $\pm$ 1.88 & 0.0286 $\pm$ 0.0017 & 14.5 $\pm$ 1.71  \\
$\lambda$7065 & 0.0395 $\pm$ 0.0010 & 22.9 $\pm$ 2.3 & 0.0289 $\pm$ 0.0009 & 29.7 $\pm$ 3.0  & 0.0259 $\pm$ 0.0020 & 12.3 $\pm$ 1.88 & 0.0324 $\pm$ 0.0017 & 19.1 $\pm$ 1.97  \\
$\lambda$3889 & 0.0696 $\pm$ 0.0037 & 7.9  $\pm$ 0.8 & 0.0884 $\pm$ 0.0011 & 16.9 $\pm$ 1.7  & 0.1046 $\pm$ 0.0070 & 12.8 $\pm$ 1.70 & 0.0877 $\pm$ 0.0055 & 8.5  $\pm$ 1.05  \\
$\lambda$4026 & 0.0160 $\pm$ 0.0021 & 2.0  $\pm$ 0.5 & 0.0160 $\pm$ 0.0011 & 3.3  $\pm$ 0.4  & 0.0206 $\pm$ 0.0021 & 2.7  $\pm$ 0.54 & 0.0188 $\pm$ 0.0012 & 2.0  $\pm$ 0.25  \\
\enddata
\end{deluxetable}

\clearpage
\begin{deluxetable}{lcccccc}
\tabletypesize{\footnotesize}
\rotate
\tablenum{3b}
\tablewidth{0pt}
\tablecaption{He Emission Lines and EW for IT98 ``High Quality'' Data Sample
\label{tbl-3b}}
\tablehead{
\colhead{He~I line} & \colhead{SBS 1159+545} && \colhead{Haro 29} &&
\colhead{SBS 1420+544} \\
& \colhead{I($\lambda$)/I(H$\beta$)} & \colhead{EW (\AA)}
& \colhead{I($\lambda$)/I(H$\beta$)} & \colhead{EW (\AA)}
& \colhead{I($\lambda$)/I(H$\beta$)} & \colhead{EW (\AA)}
}
\startdata
 & &  & from IT98 & \\
\hline
$\lambda$5876 & 0.101 $\pm$ 0.002 & 44.3 $\pm$ 4.43 & 0.102 $\pm$ 0.001 & 40.8 $\pm$ 4.08 & 0.101 $\pm$ 0.001 & 35.5 $\pm$ 3.55  \\
$\lambda$4471 & 0.039 $\pm$ 0.001 & 7.2 $\pm$ 0.72  & 0.036 $\pm$ 0.001 & 7.48 $\pm$ 0.75 & 0.037 $\pm$ 0.001 & 6.9 $\pm$ 0.69  \\
$\lambda$6678 & 0.026 $\pm$ 0.001 & 14.3 $\pm$ 1.43 & 0.029 $\pm$ 0.001 & 14.2 $\pm$ 1.42 & 0.028 $\pm$ 0.001 & 12.8 $\pm$ 1.28    \\
$\lambda$7065 & 0.027 $\pm$ 0.001 & 16.6 $\pm$ 1.66 & 0.024 $\pm$ 0.001 & 14.3 $\pm$ 1.43  & 0.037 $\pm$ 0.001 & 20.4 $\pm$ 2.04  \\
$\lambda$3889 & 0.104 $\pm$ 0.005 & 12.5 $\pm$ 1.25 & 0.097 $\pm$ 0.001 & 12.4 $\pm$ 1.24 & 0.085 $\pm$ 0.004 & 11.4 $\pm$ 1.14  \\
$\lambda$4026 & 0.019 $\pm$ 0.001 & 2.60 $\pm$ 0.26 & 0.016 $\pm$ 0.001 & 2.42 $\pm$ 0.24 & 0.018 $\pm$ 0.001 & 2.60 $\pm$ 0.26  \\
\hline
 & &  & (re-analyzed) & \\
\hline
$\lambda$5876 & 0.1018 $\pm$ 0.0021 & 44.3 $\pm$ 4.43 & 0.1017 $\pm$ 0.0010 & 40.8 $\pm$ 4.08 & 0.1006 $\pm$ 0.0019 & 35.5 $\pm$ 3.55  \\
$\lambda$4471 & 0.0386 $\pm$ 0.0010 & 7.2  $\pm$ 0.72 & 0.0366 $\pm$ 0.0010 & 7.5  $\pm$ 0.75 & 0.0376 $\pm$ 0.0011 & 6.9  $\pm$ 0.69  \\
$\lambda$6678 & 0.0267 $\pm$ 0.0010 & 14.3 $\pm$ 1.43 & 0.0286 $\pm$ 0.0010 & 14.2 $\pm$ 1.42 & 0.0283 $\pm$ 0.0009 & 12.8 $\pm$ 1.28  \\
$\lambda$7065 & 0.0274 $\pm$ 0.0010 & 16.6 $\pm$ 1.66 & 0.0247 $\pm$ 0.0010 & 14.3 $\pm$ 1.43 & 0.0373 $\pm$ 0.0010 & 20.4 $\pm$ 2.04   \\
$\lambda$3889 & 0.1033 $\pm$ 0.0047 & 12.3 $\pm$ 1.23 & 0.0969 $\pm$ 0.0011 & 12.3 $\pm$ 1.24 & 0.0893 $\pm$ 0.0037 & 11.4 $\pm$ 1.14  \\
$\lambda$4026 & 0.0197 $\pm$ 0.0011 & 2.60 $\pm$ 0.28 & 0.0158 $\pm$ 0.0010 & 2.42 $\pm$ 0.30 & 0.0187 $\pm$ 0.0011 & 2.60 $\pm$ 0.31   \\
\enddata
\end{deluxetable}

\clearpage

\begin{deluxetable}{lcccccc}
\tabletypesize{\footnotesize}
\tablenum{4}
\tablewidth{0pt}
\tablecaption{Analysis of SBS 0335-052 from IT98 \label{tbl-4}}
\tablehead{
\colhead{Analysis} &
\colhead{He$^+$/H$^+$}     &
\colhead{T$_e$ $<(He II)>$}     &
\colhead{N$_e$}     &
\colhead{He~I ABS EW (\AA)}     &
\colhead{$\tau$(3889)}     &
\colhead{$\chi ^{2}$}}
\startdata
IT98     & 0.080  $\pm$ 0.001 & 20,300 $\pm$ 300 & 67 $\pm$ 3  & 0  &
1.7 $\pm$ 0.3 & ... \\
         & (solved)  & (solved)  & (solved +)   & (fixed) & (fixed) \\
\hline
D        & 0.0798 $\pm$ 0.0016 & 20,300 $\pm$ 300 & 0.01 $^{+19}_{-0.01}$ &
            0                      & 5.06 $^{+0.32}_{-0.31}$ & 9.9  \\
D (re-an)& 0.0808 $\pm$ 0.0011 & 20,530 $\pm$ 390 & 0.01 $^{+23}_{-0}$ &
            0                      & 4.98 $^{+0.31}_{-0.30}$ & 4.4  \\
     & (solved)    & (fixed)    & (solved +)      & (fixed)  & (solved $+$) \\
\hline
D        & 0.0825 $\pm$ 0.0016 & 20,300 $\pm$ 300 & 0.02 $^{+17}_{-0.02}$ &
           0.48 $^{+0.33}_{-0.27}$ & 4.86 $^{+0.32}_{-0.31}$ & 6.8 \\
MC       & 0.0827 $\pm$ 0.0022 & 20,300 $\pm$ 300 & 0.3 $^{+8}_{-0.3}$ &
           0.52 $\pm$ 0.34     & 4.85 $\pm$ 0.33 & ... \\
D (re-an)& 0.0829 $\pm$ 0.0014 & 20,530 $\pm$ 390 & 0.0 $^{+21}_{-0}$ &
           0.38 $^{+0.33}_{-0.26}$ & 4.82 $\pm$ 0.30 & 2.3   \\
MC (re-an)& 0.0832 $\pm$ 0.0022 & 20,530 $\pm$ 390 & 3 $^{+22}_{-3}$ &
           0.44 $\pm$ 0.34 & 4.78 $\pm$0.33 & ...   \\
     & (solved)    & (fixed)    & (solved +)  & (solved $+$) & (solved $+$) \\
\hline
D        & 0.0720 $\pm$ 0.0013 & 13,990 $\pm$ 660  & 283 $^{+80}_{-88}$ &
           0.00 $^{+0.22}_{-0.00}$ & 5.22 $^{+0.35}_{-0.34}$  & 0.3  \\
MC       & 0.0735 $\pm$ 0.0042 & 14,090 $\pm$ 2,210 & 520 $^{+1138}_{-520}$ &
           0.11 $^{+0.25}_{-0.11}$  & 5.04 $\pm$ 0.83 & ... \\
D (re-an)& 0.0780 $\pm$ 0.0016 & 16,800 $\pm$ 1,120   & 63 $^{+70}_{-63}$ &
           0.13 $^{+0.26}_{-0.13}$ & 4.96 $\pm$ 0.31 & 0.1  \\
MC (re-an)& 0.0763 $\pm$ 0.0049 & 15,940 $\pm$ 2,710   & 347 $^{+941}_{-347}$ &
           0.14 $^{+0.27}_{-0.14}$ & 4.62 $\pm$0.86  & ...   \\
     & (solved)    & (solved)   & (solved +)   & (solved $+$) & (solved $+$) \\
\hline
D        & 0.0914 $\pm$ 0.0015 & 32,910 $\pm$ 500 & 0.02 $^{+8}_{-0.02}$ &
          0.42$^{+0.28}_{-0.23}$  & 8.59 $\pm$ 0.39 & 0.3 \\
MC       & 0.0712 $\pm$ 0.0055 & 13,290 $\pm$ 2,220 & 874 $^{1242}_{874}$ &
         $-$0.01$\pm$0.28 & 4.89 $\pm$ 0.90 & ... \\
D (re-an)& 0.0780 $\pm$ 0.0016 & 16,800 $\pm$ 1,120 & 63 $^{+70}_{-63}$ &
           0.13$^{+0.26}_{-0.21}$  & 4.96 $\pm$ 0.31 & 0.1 \\
MC (re-an)& 0.0755 $\pm$ 0.0065 & 15,390 $\pm$ 3,190 & 581 $^{+1273}_{-581}$ &
           0.08$\pm$0.31 & 4.50 $\pm$ 1.03 & ... \\
     & (solved)    & (solved)  & (solved $+$)  & (solved free) & (solved $+$) \\
\hline
D        & 0.0716 $\pm$ 0.0013 & 13,850 $\pm$ 650 & 307 $^{+82}_{-86}$ &
          $-$0.02$^{+0.23}_{-0.20}$  & 5.20 $\pm$ 0.35 & 0.3 \\
MC       & 0.0742 $\pm$ 0.0072 & 14,450 $\pm$ 3,170 & 348 $\pm$ 438  &
          0.07$\pm$0.31 & 5.28 $\pm$ 1.00 & ... \\
D (re-an)& 0.0780 $\pm$ 0.0016 & 16,800 $\pm$ 1,120 & 63 $^{+70}_{-65}$ &
          0.13$^{+0.26}_{-0.22}$  & 4.95 $\pm$ 0.31 & 0.1 \\
MC (re-an)& 0.0762 $\pm$ 0.0098 & 15,730 $\pm$ 4,700 & 771 $\pm$ 1,040 &
          0.09$\pm$0.38 & 4.57 $\pm$ 1.38 & ... \\
     & (solved)   & (solved)  & (solved free)  & (solved free) & (solved free) \\
\enddata
\end{deluxetable}

\newpage

\begin{deluxetable}{lcccccc}
\tabletypesize{\footnotesize}
\tablenum{5}
\tablewidth{0pt}
\tablecaption{Analysis of NGC 2363A from IT98 \label{tbl-5}}
\tablehead{
\colhead{Analysis} &
\colhead{He$^+$/H$^+$}     &
\colhead{T$_e$ $<(He II)>$}     &
\colhead{N$_e$}     &
\colhead{He~I ABS EW (\AA)}     &
\colhead{$\tau$(3889)}     &
\colhead{$\chi ^{2}$}}
\startdata
IT98     & 0.081  $\pm$ 0.001 & 15,800 $\pm$ 100 & 253 $\pm$ 10  & 0  &
0.0 & ... \\
         & (solved)  & (solved)  & (solved +)   & (fixed) & (fixed) \\
\hline
D        & 0.0840 $\pm$ 0.0010 & 15,800 $\pm$ 100 & 0.4 $^{+40}_{-0.4}$ &
            0                      & 2.19 $^{+0.16}_{-0.17}$ & 16.5  \\
D (re-an)& 0.0840 $\pm$ 0.0010 & 15,870 $\pm$ 60 & 0.1 $^{+42}_{-0.1}$ &
            0                      & 2.06 $^{+0.17}_{-0.17}$ & 15.2  \\
     & (solved)    & (fixed)    & (solved +)      & (fixed)  & (solved $+$) \\
\hline
D        & 0.0865 $\pm$ 0.0007 & 15,800 $\pm$ 100 & 0.01 $^{+22}_{-0.01}$ &
           0.75 $^{+0.23}_{-0.21}$ & 1.98 $^{+0.17}_{-0.18}$ & 2.5 \\
MC       & 0.0865 $\pm$ 0.0010 & 15,800 $\pm$ 100 & 2.5 $^{+18}_{-2.5}$  &
           0.77 $\pm$ 0.24 & 1.97 $\pm$ 0.19 & ... \\
D (re-an)& 0.0862 $\pm$ 0.0008 & 15,870 $\pm$  60 & 0.02 $^{+26}_{-0.02}$ &
           0.73 $^{+0.24}_{-0.22}$ & 1.87 $^{+0.17}_{-0.18}$ & 2.7   \\
MC (re-an)& 0.0861 $\pm$ 0.0011 & 15,870 $\pm$  60 & 6 $^{+29}_{-6}$ &
           0.75 $\pm$ 0.25 & 1.85 $\pm$ 0.19 & ...   \\
     & (solved)    & (fixed)    & (solved +)  & (solved $+$) & (solved $+$) \\
\hline
D        & 0.0848 $\pm$ 0.0009 & 14,720 $\pm$  510  & 28 $^{+67}_{-28}$ &
           0.57 $^{+0.21}_{-0.19}$ & 1.84 $\pm$ 0.17  & 0.4  \\
MC       & 0.0824 $\pm$ 0.0042 & 14,260 $\pm$ 1,030 & 176 $^{+402}_{-176}$  &
           0.46 $\pm$ 0.28     & 1.66 $\pm$ 0.38 & ... \\
D (re-an)& 0.0803 $\pm$ 0.0009 & 14,010 $\pm$ 460   & 205 $^{+80}_{-74}$ &
           0.35 $^{+0.21}_{-0.19}$ & 1.45 $\pm$ 0.18 & 0.8  \\
MC (re-an)& 0.0801 $\pm$ 0.0046 & 14,040 $\pm$ 1,060  & 285 $^{+343}_{-285}$ &
           0.35 $\pm$ 0.28 & 1.44 $\pm$ 0.40 & ...   \\
     & (solved)    & (solved)   & (solved +)   & (solved $+$) & (solved $+$) \\
\hline
D        & 0.0847 $\pm$ 0.0009 & 14,720 $\pm$  510  & 28 $^{+65}_{-28}$ &
            0.57 $^{+0.21}_{-0.19}$  & 1.84 $\pm$ 0.17 & 0.4 \\
MC       & 0.0823 $\pm$ 0.0046 & 14,240 $\pm$ 1,080 & 202 $^{+512}_{-202}$ &
            0.45 $\pm$ 0.31 & 1.65 $\pm$ 0.40 & ... \\
D (re-an)& 0.0804 $\pm$ 0.0009 & 14,150  $\pm$   460  & 204 $^{+78}_{-73}$ &
            0.35$^{+0.21}_{-0.19}$  & 1.45 $\pm$ 0.18 & 0.8 \\
MC (re-an)& 0.0797 $\pm$ 0.0054 & 13,970 $\pm$ 1,160 & 346 $^{+504}_{-346}$ &
            0.32$\pm$0.34 & 1.41 $\pm$ 0.43 & ... \\
     & (solved)    & (solved)  & (solved $+$)  & (solved free) & (solved $+$) \\
\hline
D        & 0.0847 $\pm$ 0.0009 & 14,710 $\pm$ 510 & 29 $^{+65}_{-61}$ &
           0.56 $^{+0.21}_{-0.19}$  & 1.84 $\pm$ 0.17 & 0.4 \\
MC       & 0.0885 $\pm$ 0.0105 & 16,440 $\pm$ 3,690 &  91 $\pm$ 406 &
           0.72 $\pm$0.52 & 2.29 $\pm$ 1.09 & ... \\
D (re-an)& 0.0804 $\pm$ 0.0009 & 14,020 $\pm$  460  & 203 $^{+79}_{-73}$ &
           0.35$^{+0.21}_{-0.19}$  & 1.45 $\pm$ 0.18 & 0.8 \\
MC (re-an)& 0.0819 $\pm$ 0.0089 & 14,910 $\pm$ 3,300 & 311 $\pm$ 448 &
           0.42$\pm$0.46 & 1.67 $\pm$ 0.94 & ... \\
     & (solved)   & (solved)  & (solved free)  & (solved free) & (solved free) \\
\enddata
\end{deluxetable}
\newpage

\begin{deluxetable}{lcccccc}
\tabletypesize{\footnotesize}
\tablenum{6}
\tablewidth{0pt}
\tablecaption{Analysis of SBS 0940+544 from IT98 \label{tbl-6}}
\tablehead{
\colhead{Analysis} &
\colhead{He$^+$/H$^+$}     &
\colhead{T$_e$ $<(He II)>$}     &
\colhead{N$_e$}     &
\colhead{He~I ABS EW (\AA)}     &
\colhead{$\tau$(3889)}     &
\colhead{$\chi ^{2}$}}
\startdata
IT98     & 0.083  $\pm$ 0.002 & 20,200 $\pm$ 300 & 10  & 0  &
0.0 & ... \\
         & (solved)  & (solved)  & (solved +)   & (fixed) & (fixed) \\
\hline
D        & 0.0818 $\pm$ 0.0016 & 20,200 $\pm$ 300 & 72 $^{+73}_{-63}$ &
            0                      & 0.0 $^{+0.37}_{-0.00}$ & 4.0  \\
D (re-an)& 0.0852 $\pm$ 0.0019 & 20,200 $\pm$ 400 & 0.3 $^{+68}_{-0.3}$ &
            0                      & 0.35 $^{+0.51}_{-0.35}$ & 1.5  \\
     & (solved)    & (fixed)    & (solved +)      & (fixed)  & (solved $+$) \\
\hline
D        & 0.0817 $\pm$ 0.0018 & 20,200 $\pm$ 300 & 75 $^{+69}_{-68}$ &
           0.00 $^{+0.14}_{-0.00}$ & 0.00 $^{+0.35}_{-0.00}$ & 4.0 \\
MC       & 0.0824 $\pm$ 0.0026 & 20,200 $\pm$ 300 & 67 $\pm$ 63 &
           0.05 $^{+0.22}_{-0.05}$  & 0.12 $^{+0.41}_{-0.12}$ & ... \\
D (re-an)& 0.0852 $\pm$ 0.0023 & 20,200 $\pm$ 400 & 0.1 $^{+68}_{-0.1}$ &
           0.00 $^{+0.24}_{-0.00}$ & 0.35 $^{+0.51}_{-0.35}$ & 1.5   \\
MC (re-an)& 0.0852 $\pm$ 0.0031 & 20,200 $\pm$ 400 & 22 $^{+61}_{-22}$ &
           0.09 $^{+0.33}_{-0.09}$ & 0.31 $^{+0.47}_{-0.31}$ & ...   \\
     & (solved)    & (fixed)    & (solved +)  & (solved $+$) & (solved $+$) \\
\hline
D        & 0.0843 $\pm$ 0.0024 & 22,920 $\pm$ 4,110 & 42 $^{+52}_{-42}$ &
           0.00 $^{+0.20}_{-0.00}$ & 0.00 $^{+0.49}_{-0.00}$  & 3.5  \\
MC       & 0.0824 $\pm$ 0.0043 & 20,330 $\pm$ 2,510 & 83 $^{+230}_{-83}$  &
           0.05 $^{+0.24}_{-0.05}$ & 0.22 $^{+0.45}_{-0.22}$ & ... \\
D (re-an)& 0.0852 $\pm$ 0.0027 & 20,190 $\pm$ 3,770   & 0.5 $^{+65}_{-0.5}$ &
           0.00 $^{+0.24}_{-0.00}$ & 0.35 $^{+0.51}_{-0.35}$  & 1.5  \\
MC (re-an)& 0.0841 $\pm$ 0.0035 & 19,260 $\pm$ 2,480  & 35 $^{+140}_{-35}$ &
           0.07 $^{+0.27}_{-0.07}$ & 0.36 $^{+0.46}_{-0.36}$  & ...   \\
     & (solved)    & (solved)   & (solved +)   & (solved $+$) & (solved $+$) \\
\hline
D        & 0.0814 $\pm$ 0.0025 & 21,640 $\pm$ 3,880 & 64 $^{+62}_{-56}$ &
          $-$0.21 $^{+0.30}_{-0.26}$  & 0.00 $^{+0.37}_{-0.00}$ & 3.1 \\
MC       & 0.0735 $\pm$ 0.0105 & 18,300 $\pm$ 3.510 & 456 $^{+587}_{-456}$ &
          $-$0.46 $\pm$ 0.45 & 0.12 $^{+0.41}_{-0.12}$ & ... \\
D (re-an)& 0.0799 $\pm$ 0.0026 & 18,330  $\pm$  2,510 & 93 $^{+103}_{-84}$ &
         $-$0.25$^{+0.29}_{-0.26}$  & 0.00 $^{+0.44}_{-0.00}$ & 1.2 \\
MC (re-an)& 0.0761 $\pm$ 0.0111 & 17,970 $\pm$ 3,030 & 354 $^{+696}_{-354}$ &
         $-$0.34$\pm$0.48 & 0.23 $^{+0.48}_{-0.23}$ & ... \\
     & (solved)    & (solved)  & (solved $+$)  & (solved free) & (solved $+$) \\
\hline
D        & 0.0705 $\pm$ 0.0021 & 17,960 $\pm$ 1,490 & 394 $^{+166}_{-132}$ &
         $-$0.58 $^{+0.24}_{-0.22}$  & $-$1.05 $\pm$ 0.49 & 1.6 \\
MC       & 0.0711 $\pm$ 0.0070 & 18,490 $\pm$ 2,560 & 434 $\pm$ 296 &
         $-$0.55 $\pm$0.33 & $-$0.73 $\pm$ 0.69 & ... \\
D (re-an)& 0.0732 $\pm$ 0.0025 & 16,870 $\pm$ 1,620 & 318 $^{+218}_{-144}$ &
         $-$0.45$^{+0.27}_{-0.25}$  & $-$0.62 $^{+0.54}_{-0.51}$ & 0.9 \\
MC (re-an)& 0.0740 $\pm$ 0.0110 & 17,840 $\pm$ 2,990 & 420 $\pm$ 454 &
         $-$0.42$\pm$0.45 & $-$0.06 $\pm$ 0.89 & ... \\
     & (solved)   & (solved)  & (solved free)  & (solved free) & (solved free) \\
\enddata
\end{deluxetable}

\newpage

\begin{deluxetable}{lcccccc}
\tabletypesize{\footnotesize}
\tablenum{7}
\tablewidth{0pt}
\tablecaption{Analysis of MRK 193 from IT98 \label{tbl-7}}
\tablehead{
\colhead{Analysis} &
\colhead{He$^+$/H$^+$}     &
\colhead{T$_e$ $<(He II)>$}     &
\colhead{N$_e$}     &
\colhead{He~I ABS EW (\AA)}     &
\colhead{$\tau$(3889)}     &
\colhead{$\chi ^{2}$}}
\startdata
IT98     & 0.081  $\pm$ 0.001 & 16,600 $\pm$ 200 & 326 $\pm$ 65  & 0  &
0.0 & ... \\
         & (solved)  & (solved)  & (solved +)   & (fixed) & (fixed) \\
\hline
D        & 0.0784 $\pm$ 0.0011 & 16,600 $\pm$ 200 & 448 $^{+154}_{-140}$ &
            0                      & 0.56 $\pm$ 0.44 & 1.6  \\
D (re-an)& 0.0866 $\pm$ 0.0016 & 16,560 $\pm$ 200 & 54 $^{+94}_{-54}$ &
            0                      & 2.16 $^{+0.44}_{-0.42}$ & 5.2  \\
     & (solved)    & (fixed)    & (solved +)      & (fixed)  & (solved $+$) \\
\hline
D        & 0.0804 $\pm$ 0.0018 & 16,600 $\pm$ 200 & 387 $^{+145}_{-127}$ &
           0.15 $^{+0.27}_{-0.15}$ & 0.62 $^{+0.44}_{-0.45}$ & 1.4 \\
MC       & 0.0813 $\pm$ 0.0049 & 16,600 $\pm$ 200 & 369 $\pm$ 188 &
           0.22 $^{+0.27}_{-0.22}$  & 0.74 $\pm$ 0.53 & ... \\
D (re-an)& 0.0896 $\pm$ 0.0019 & 16,560 $\pm$ 200 & 0.04 $^{+66}_{-0.04}$ &
           0.20 $^{+0.19}_{-0.16}$ & 2.30 $^{+0.43}_{-0.41}$ & 3.8   \\
MC (re-an)& 0.0888 $\pm$ 0.0035 & 16,560 $\pm$ 200 & 38 $^{+142}_{-38}$ &
           0.20 $\pm$ 0.18 & 2.16 $\pm$0.51 & ...   \\
     & (solved)    & (fixed)    & (solved +)  & (solved $+$) & (solved $+$) \\
\hline
D        & 0.0759 $\pm$ 0.0017 & 14,890 $\pm$ 990  & 637 $^{+199}_{-179}$ &
           0.00 $^{+0.20}_{-0.00}$ & 0.61 $^{+0.39}_{-0.38}$  & 0.4  \\
MC       & 0.0781 $\pm$ 0.0054 & 15,500 $\pm$ 1,860 & 607 $\pm$ 372 &
           0.11 $^{+0.30}_{-0.11}$ & 0.72 $\pm$ 0.49 & ... \\
D (re-an)& 0.0795 $\pm$ 0.0016 & 12,600 $\pm$ 850   & 524 $\pm$ 164 &
           0.00 $^{+0.10}_{-0.00}$ & 1.81 $^{+0.41}_{-0.42}$  & 1.0  \\
MC (re-an)& 0.0802 $\pm$ 0.0046 & 12,790 $\pm$ 1,760   & 820 $^{+860}_{-820}$ &
           0.02 $^{+0.13}_{-0.02}$ & 1.89 $\pm$0.70  & ...   \\
     & (solved)    & (solved)   & (solved +)   & (solved $+$) & (solved $+$) \\
\hline
D        & 0.0746 $\pm$ 0.0017 & 14,640 $\pm$ 970 & 716 $^{+213}_{-190}$ &
         $-$0.08 $^{+0.22}_{-0.19}$  & 0.62 $\pm$ 0.37 & 0.4 \\
MC       & 0.0758 $\pm$ 0.0071 & 15,040 $\pm$ 2,010 & 790 $\pm$ 513 &
         $-$0.02 $\pm$0.33 & 0.76 $\pm$ 0.45 & ... \\
D (re-an)& 0.0703 $\pm$ 0.0015 & 11,240  $\pm$   460 & 2,413 $^{+84}_{-108}$ &
         $-$0.27 $\pm$ 0.13  & 1.55 $\pm$ 0.15 & 0.1 \\
MC (re-an)& 0.0717 $\pm$ 0.0070 & 11,900 $\pm$ 1,300 & 1,925 $\pm$ 866 &
         $-$0.24$\pm$0.22 & 1.86 $\pm$ 0.80 & ... \\
     & (solved)    & (solved)  & (solved $+$)  & (solved free) & (solved $+$) \\
\hline
D        & 0.0746 $\pm$ 0.0017 & 14,630 $\pm$ 970 & 718 $^{+213}_{-191}$ &
         $-$0.08$^{+0.22}_{-0.19}$  & 0.62 $\pm$ 0.37 & 0.4 \\
MC       & 0.0752 $\pm$ 0.0068 & 14,980 $\pm$ 1,870 & 822 $\pm$ 511 &
         $-$0.04$\pm$0.32 & 0.73 $\pm$ 0.46 & ... \\
D (re-an)& 0.0703 $\pm$ 0.0015 & 11,240 $\pm$ 460 & 2,413 $^{+84}_{-108}$ &
         $-$0.27 $\pm$ 0.13  & 1.55 $\pm$ 0.15 & 0.1 \\
MC (re-an)& 0.0719 $\pm$ 0.0075 & 11,940 $\pm$ 1,440 & 1,916 $\pm$ 885 &
         $-$0.23$\pm$0.22 & 1.89 $\pm$ 0.85 & ... \\
     & (solved)   & (solved)  & (solved free)  & (solved free) & (solved free) \\
\enddata
\end{deluxetable}

\newpage

\begin{deluxetable}{lcccccc}
\tabletypesize{\footnotesize}
\tablenum{8}
\tablewidth{0pt}
\tablecaption{Analysis of SBS 1159+545 from IT98 \label{tbl-8}}
\tablehead{
\colhead{Analysis} &
\colhead{He$^+$/H$^+$}     &
\colhead{T$_e$ $<(He II)>$}     &
\colhead{N$_e$}     &
\colhead{He~I ABS EW (\AA)}     &
\colhead{$\tau$(3889)}     &
\colhead{$\chi ^{2}$}}
\startdata
IT98     & 0.081  $\pm$ 0.002 & 18,400 $\pm$ 200 & 110 $\pm$ 58  & 0  & 0 & ... \\
         & (solved)   & (solved)  & (solved +)      & (fixed) & (fixed) \\
\hline
D        & 0.0811 $\pm$ 0.0010 & 18,400 $\pm$ 200 & 130 $^{+53}_{-47}$ &
            0                      & 0.00 $^{+0.28}_{-0.00}$ & 3.5  \\   
D (re-an)& 0.0828 $\pm$ 0.0011 & 18,570 $\pm$ 210 & 83 $^{+47}_{-44}$ &
            0                      & 0.31 $^{+0.29}_{-0.28}$ & 3.3  \\
     & (solved)    & (fixed)     & (solved +)    & fixed  & (solved $+$) \\
\hline
D        & 0.0811 $\pm$ 0.0011 & 18,400 $\pm$ 200 & 129 $^{+52}_{-46}$ &
           0.00 $^{+0.06}_{-0.00}$ & 0.00 $^{+0.28}_{-0.00}$ & 3.5  \\   
MC       & 0.0820 $\pm$ 0.0020 & 18,400 $\pm$ 200 & 100 $\pm$ 62 &
           0.01$^{+0.07}_{-0.01}$  & 0.22 $^{+0.39}_{-0.22}$ & ... \\
D (re-an)& 0.0828 $\pm$ 0.0012 & 18,570 $\pm$ 210 & 83 $^{+45}_{-44}$ &
           0.00 $^{+0.07}_{-0.00}$ & 0.32 $^{+0.28}_{-0.29}$ & 3.3  \\
MC (re-an)& 0.0832 $\pm$ 0.0021 & 18,570 $\pm$ 210 & 76 $\pm$ 62 &
           0.01 $^{+0.08}_{-0.01}$  & 0.39 $\pm$ 0.37 & ... \\
     & (solved)    & (fixed)     & (solved +)   & (solved $+$) & (solved $+$) \\
\hline
D        & 0.0825 $\pm$ 0.0013 & 19,460 $\pm$ 1,480 & 93 $^{+44}_{-39}$ &
           0.00 $^{+0.08}_{-0.00}$ & 0.05 $^{+0.28}_{-0.05}$  & 3.3  \\  
MC       & 0.0830 $\pm$ 0.0033 & 19,420 $\pm$ 2,100 & 89 $^{+101}_{-89}$  &
           0.02 $^{+0.10}_{-0.02}$  & 0.27 $^{+0.39}_{-0.27}$  & ... \\
D (re-an)& 0.0840 $\pm$ 0.0014 & 19,390 $\pm$ 1,670 & 55 $^{+41}_{-38}$ &
           0.00 $^{+0.09}_{-0.00}$ & 0.38 $^{+0.28}_{-0.27}$  & 3.2  \\
MC (re-an)& 0.0838 $\pm$ 0.0031 & 19,330 $\pm$ 2,100 & 75 $^{+116}_{-75}$ &
           0.01 $^{+0.07}_{-0.01}$ & 0.45 $\pm$ 0.38  & ...   \\
     & (solved)    & (solved)    & (solved +)   & (solved $+$) & (solved $+$) \\
\hline
D        & 0.0748 $\pm$ 0.0013 & 16,610 $\pm$ 860 & 297 $^{+93}_{-77}$ &
           $-$0.32 $^{+0.14}_{-0.13}$  & 0.00 $^{+0.15}_{-0.00}$ & 1.7 \\ 
MC       & 0.0725 $\pm$ 0.0111 & 17,550 $\pm$ 2,720 & 529 $\pm$ 484 &
           $-$0.47 $\pm$ 0.46  & 0.15 $^{+0.43}_{-0.15}$ & ... \\   
D (re-an)& 0.0741 $\pm$ 0.0013 & 16,280 $\pm$ 800 & 342 $^{+102}_{-86}$ &
           $-$0.38 $^{+0.14}_{-0.13}$  & 0.00 $^{+0.19}_{-0.00}$ & 1.0  \\
MC (re-an)& 0.0742 $\pm$ 0.0080 & 16,980 $\pm$ 2,330 & 434 $\pm$ 378 &
           $-$0.39 $\pm$ 0.27 & 0.17 $^{+0.43}_{-0.17}$ & ...  \\
     & (solved)    & (solved)   & (solved +)    & (solved free) & (solved $+$) \\
\hline
D        & 0.0697 $\pm$ 0.0013 & 16,570 $\pm$ 780 & 537 $^{+143}_{-126}$ &
           $-$0.48 $\pm$ 0.13  & $-$0.52 $\pm$ 0.24 & 0.01 \\   
MC       & 0.0698 $\pm$ 0.0058 & 16,970 $\pm$ 1,650 & 564 $\pm$ 276 &
           $-$0.48 $\pm$ 0.22  & $-$0.38 $\pm$ 0.44 & ...  \\   
D (re-an)& 0.0716 $\pm$ 0.0014 & 16,280 $\pm$ 760 & 466 $^{+138}_{-109}$ &
           $-$0.46 $^{+0.14}_{-0.13}$  & $-$0.31 $\pm$ 0.26 & 0.45 \\
MC (re-an)& 0.0720 $\pm$ 0.0062 & 16,710 $\pm$ 1,840 & 493 $\pm$ 287 &
           $-$0.45 $\pm$ 0.22  & $-$0.13 $\pm$ 0.47 & ...  \\
     & (solved)    & (solved)  & (solved free)  & (solved free) & (solved free) \\
\enddata
\end{deluxetable}

\newpage

\begin{deluxetable}{lcccccc}
\tabletypesize{\footnotesize}
\tablenum{9}
\tablewidth{0pt}
\tablecaption{Analysis of Haro 29 (SBS 1223+487) from IT98
\label{tbl-9}}
\tablehead{
\colhead{Analysis} &
\colhead{He$^+$/H$^+$}     &
\colhead{T$_e$ $<(He II)>$}     &
\colhead{N$_e$}     &
\colhead{He~I ABS EW (\AA)}     &
\colhead{$\tau$(3889)}     &
\colhead{$\chi ^{2}$}}
\startdata
IT98     & 0.083  $\pm$ 0.001 & 16,180 $\pm$ 100 & 11 $\pm$ 2  & 0  & 0 & ... \\
         & (solved)   & (solved)  & (solved)      & (fixed) & (fixed) \\
\hline
D        & 0.0801 $\pm$ 0.0009 & 16,180 $\pm$ 100 & 92 $^{+63}_{-58}$ &
            0                      & 0.20 $\pm$ 0.18 & 13.4  \\
D (re-an)& 0.0789 $\pm$ 0.0007 & 16,150 $\pm$ 60 & 144 $^{+70}_{-63}$ &
            0                      & 0.16 $^{+0.18}_{-0.16}$ & 9.2  \\
     & (solved)    & (fixed)   & (solved +)      & fixed  & (solved $+$) \\
\hline
D        & 0.0841 $\pm$ 0.0009 & 16,180 $\pm$ 100 & 26 $^{+54}_{-26}$ &
           0.50 $^{+0.17}_{-0.15}$ & 0.20 $\pm$ 0.18 & 2.4  \\
MC       & 0.0839 $\pm$ 0.0017 & 16,180 $\pm$ 100 & 38 $^{+54}_{-38}$ &
           0.50 $\pm$ 0.17     & 0.20 $\pm$ 0.17 & ... \\
D (re-an)& 0.0827 $\pm$ 0.0009 & 16,150 $\pm$ 60 & 69 $^{+57}_{-54}$ &
           0.42 $^{+0.16}_{-0.15}$ & 0.20 $\pm$ 0.19 & 1.2  \\
MC (re-an)& 0.0827 $\pm$ 0.0020 & 16,150 $\pm$ 60 & 71 $\pm$ 60 &
           0.42 $\pm$ 0.16 & 0.22 $\pm$ 0.19 & ...  \\
     & (solved)    & (fixed)    & (solved +)  & (solved $+$) & (solved $+$) \\
\hline
D        & 0.0853 $\pm$ 0.0010 & 16,560 $\pm$ 600   & 0.02 $^{+44}_{-0.02}$ &
           0.55 $^{+0.17}_{-0.16}$ & 0.31 $\pm$ 0.18  & 2.3  \\
MC       & 0.0847 $\pm$ 0.0019 & 16,470 $\pm$ 780 & 19 $^{+75}_{-19}$ &
           0.53 $\pm$ 0.20     & 0.27 $\pm$ 0.21 & ... \\
D (re-an)& 0.0854 $\pm$ 0.0010 & 16,960 $\pm$ 640  & 0.1 $^{+44}_{-0.1}$ &
           0.53 $^{+0.17}_{-0.16}$ & 0.44 $^{+0.18}_{-0.19}$  & 0.8  \\
MC (re-an)& 0.0844 $\pm$ 0.0024 & 16,760 $\pm$ 880  & 31 $^{+101}_{-31}$ &
           0.50 $\pm$ 0.20 & 0.37 $\pm$ 0.24  & ...  \\
     & (solved)    & (solved)  & (solved +)   & (solved $+$) & (solved $+$) \\
\hline
D        & 0.0853 $\pm$ 0.0010 & 16,560 $\pm$ 600 & 0.1 $^{+44}_{-0.1}$ &
           0.55 $^{+0.17}_{-0.16}$  & 0.31 $\pm$ 0.18 & 2.3 \\
MC       & 0.0848 $\pm$ 0.0018 & 16,510 $\pm$ 760 & 14 $^{+70}_{-14}$ &
           0.54 $\pm$ 0.20 & 0.28 $\pm$ 0.21 & ... \\
D (re-an)& 0.0854 $\pm$ 0.0010 & 16,960 $\pm$ 640 & 0.1 $^{+43}_{-0.1}$ &
           0.53 $^{+0.17}_{-0.16}$  & 0.44 $\pm$ 0.19 & 0.8 \\
MC (re-an)& 0.0846 $\pm$ 0.0022 & 16,810 $\pm$ 840 & 25 $^{+97}_{-25}$ &
           0.50 $\pm$ 0.19  & 0.38 $\pm$ 0.24 & ... \\
     & (solved)  & (solved) & (solved $+$)  & (solved free) & (solved $+$) \\
\hline
D        & 0.0922 $\pm$ 0.0011 & 18,650 $\pm$ 700 & $-$125 $^{+31}_{-29}$ &
           0.81 $^{+0.19}_{-0.18}$  & 0.98  $\pm$ 0.18 & 1.9 \\
MC       & 0.0905 $\pm$ 0.0068 & 18,430 $\pm$ 2,190 & $-$60 $\pm$ 151 &
           0.74 $\pm$ 0.32  & 0.88 $\pm$ 0.69 & ...  \\
D (re-an)& 0.0909 $\pm$ 0.0011 & 18,620 $\pm$ 760 & $-$100 $^{+33}_{-31}$ &
           0.73 $^{+0.19}_{-0.17}$  & 0.98 $\pm$ 0.18 & 0.6 \\
MC (re-an)& 0.0896 $\pm$ 0.0075 & 18,680 $\pm$ 2,440 & $-$32 $\pm$ 175 &
           0.67 $\pm$ 0.33  & 0.96 $\pm$ 0.75 & ... \\
     & (solved) & (solved)  & (solved free)  & (solved free) & (solved free) \\
\enddata
\end{deluxetable}

\newpage

\begin{deluxetable}{lcccccc}
\tabletypesize{\footnotesize}
\tablenum{10}
\tablewidth{0pt}
\tablecaption{Analysis of SBS 1420+544 from IT98 \label{tbl-10}}
\tablehead{
\colhead{Analysis} &
\colhead{He$^+$/H$^+$}     &
\colhead{T$_e$ $<(He II)>$}     &
\colhead{N$_e$}     &
\colhead{He~I ABS EW (\AA)}     &
\colhead{$\tau$(3889)}     &
\colhead{$\chi ^{2}$}}
\startdata
IT98     & 0.082  $\pm$ 0.002 & 17,600 $\pm$ 100 &  26 $\pm$ 7  & 0  & 1.8 $\pm$ 0.3 & ... \\
         & (solved)   & (solved)   & (solved +)      & (fixed) & (solved) \\
\hline
D        & 0.0797 $\pm$ 0.0007 & 17,600 $\pm$ 100 & 128 $^{+56}_{-52}$ &
            0                      & 3.17 $\pm$ 0.28 & 2.3  \\
D (re-an)& 0.0808 $\pm$ 0.0009 & 17,850 $\pm$ 120 & 119 $^{+52}_{-48}$ &
            0                      & 3.04 $\pm$ 0.27 & 3.6   \\
     & (solved)    & (fixed)   & (solved +)      & fixed  & (solved $+$) \\
\hline
D        & 0.0807 $\pm$ 0.0010 & 17,600 $\pm$ 100 & 98 $^{+56}_{-52}$ &
           0.07 $^{+0.15}_{-0.07}$ & 3.28 $\pm$ 0.28 & 2.2  \\
MC       & 0.0811 $\pm$ 0.0025 & 17,600 $\pm$ 100 & 92 $\pm$ 83 &
           0.11 $\pm$ 0.12     & 3.32 $\pm$ 0.44 & ... \\
D (re-an)& 0.0808 $\pm$ 0.0011 & 17,850 $\pm$ 120 & 120 $^{+52}_{-50}$ &
           0.00 $^{+0.12}_{-0.00}$ & 3.04 $\pm$ 0.27 & 3.6   \\
MC (re-an)& 0.0815 $\pm$ 0.0024 & 17,850 $\pm$ 120 & 110 $\pm$ 79 &
           0.04 $^{+0.14}_{-0.04}$ & 3.08 $\pm$ 0.43 & ...   \\
     & (solved)    & (fixed) & (solved +)   & (solved $+$) & (solved $+$) \\
\hline
D        & 0.0852 $\pm$ 0.0013 & 19,160 $\pm$ 1,710 & 2 $^{+40}_{-2}$ &
           0.23 $^{+0.17}_{-0.15}$ & 3.54 $\pm$ 0.28 & 2.0  \\
MC       & 0.0831 $\pm$ 0.0035 & 18,680 $\pm$ 1,700  & 56 $^{+116}_{-56}$ &
           0.16 $\pm$ 0.17     & 3.43 $\pm$ 0.39 & ... \\
D (re-an)& 0.0866 $\pm$ 0.0015 & 20,940 $\pm$ 2,290 & 0.3 $^{+30}_{-0.3}$ &
           0.13 $^{+0.19}_{-0.13}$ & 3.41 $\pm$ 0.27 & 2.3   \\
MC (re-an)& 0.0854 $\pm$ 0.0026 & 20,370 $\pm$ 1,740 & 27 $^{+83}_{-27}$ &
           0.12 $^{+0.18}_{-0.12}$ & 3.32 $\pm$ 0.35  & ...   \\
     & (solved)    & (solved)  & (solved +)   & (solved $+$) & (solved $+$) \\
\hline
D        & 0.0852 $\pm$ 0.0013 & 19,130 $\pm$ 1,710 & 4 $^{+39}_{-4}$ &
           0.22 $^{+0.16}_{-0.15}$  & 3.52 $\pm$ 0.28  & 2.0  \\
MC       & 0.0804 $\pm$ 0.0077 & 17,280 $\pm$ 3,640 & 419 $^{+1,039}_{-419}$ &
           0.06 $\pm$ 0.30 & 3.11 $\pm$ 0.74 & ...  \\
D (re-an)& 0.0866 $\pm$ 0.0015 & 20,950 $\pm$ 2,290 & 0.3 $^{+30}_{-0.3}$ &
           0.13 $^{+0.19}_{-0.16}$ & 3.41 $\pm$ 0.27 & 2.3  \\
MC (re-an)& 0.0860 $\pm$ 0.0030 & 20,600 $\pm$ 2,070 & 25 $^{+359}_{-25}$ &
           0.11 $\pm$ 0.20  & 3.44 $\pm$ 0.34 & ...  \\
     & (solved)    & (solved)    & (solved +)  & (solved free) & (solved $+$) \\
\hline
D        & 0.0854 $\pm$ 0.0012 & 19,200 $\pm$ 1,720 & $-$1 $^{+18}_{-18}$ &
           0.23 $^{+0.16}_{-0.15}$  & 3.55 $\pm$ 0.28 & 1.96 \\
MC (re-an)& 0.0802 $\pm$ 0.0091 & 17,340 $\pm$ 3,920 & 459 $\pm$ 746 &
           0.06 $\pm$ 0.34  & 3.14 $\pm$ 0.92 & ...  \\
D (re-an)& 0.0870 $\pm$ 0.0016 & 21,150 $\pm$ 2,240  & $-$6 $^{+29}_{-28}$  &
           0.14 $^{+0.19}_{-0.16}$ & 3.46 $^{+0.27}_{-0.26}$ & 2.3  \\
MC (re-an)& 0.0855 $\pm$ 0.0057 & 20,450 $\pm$ 2,400 & 72 $\pm$ 309 &
           0.10 $\pm$ 0.26  & 3.41 $\pm$ 0.58 & ...  \\
     & (solved)   & (solved) & (solved free)  & (solved free) & (solved free) \\
\enddata
\end{deluxetable}

\newpage

\begin{deluxetable}{lcccccccc}
\tabletypesize{\footnotesize}
\rotate
\tablenum{11}
\tablewidth{0pt}
\tablecaption{Summary of Re-Analysis of IT98 ``High Quality'' Data Sample \label{tbl-11}}
\tablehead{
\colhead{Name} &
\colhead{T (K)} &
\colhead{n (cm$^{-3}$)} &
\colhead{$\tau$(3889)} &
\colhead{He(ABS) (EW)} &
\colhead{He$^+$/H$^+$} &
\colhead{He$^{++}$/H$^+$} &
\colhead{Y} &
\colhead{O/H $\times$ 10$^4$}
}
\startdata
&&&& IT98 \\
\hline
SBS 0335-052 & 20,300 $\pm$ 300 & 67 $\pm$ 3   & 1.7 $\pm$ 0.3 & 0 & 
0.080 $\pm$ 0.001 & 0.002 $\pm$ 0.001 & 0.2488 $\pm$ 0.0044 & 0.19 $\pm$ 0.01 \\ 
NGC 2363A    & 15,800 $\pm$ 100 & 253 $\pm$ 10 & 0.0           & 0 & 
0.081 $\pm$ 0.001 & 0.001 $\pm$ 0.001 & 0.2456 $\pm$ 0.0008 & 0.71 $\pm$ 0.01 \\
SBS 0940+544N& 20,200 $\pm$ 300 &  10          & 0.0           & 0 & 
0.082 $\pm$ 0.002 & 0.000             & 0.2500 $\pm$ 0.0057 & 0.27 $\pm$ 0.01 \\ 
MRK 193      & 16,600 $\pm$ 200 & 326 $\pm$ 65 & 0.0           & 0 & 
0.081 $\pm$ 0.001 & 0.001 $\pm$ 0.001 & 0.2478 $\pm$ 0.0037 & 0.64 $\pm$ 0.01 \\ 
SBS 1159+545 & 18,400 $\pm$ 200 & 110 $\pm$ 58 & 0.0           & 0 & 
0.081 $\pm$ 0.002 & 0.001 $\pm$ 0.001 & 0.2456 $\pm$ 0.0049 & 0.31 $\pm$ 0.01 \\ 
HARO 29      & 16,180 $\pm$ 100 &  11 $\pm$  2 & 0.0           & 0 & 
0.083 $\pm$ 0.001 & 0.001 $\pm$ 0.001 & 0.2509 $\pm$ 0.0012 & 0.59 $\pm$ 0.01 \\ 
SBS 1420+544 & 17,600 $\pm$ 100 &  26 $\pm$  7 & 1.8 $\pm$ 0.3 & 0 & 
0.082 $\pm$ 0.001 & 0.001 $\pm$ 0.001 & 0.2497 $\pm$ 0.0030 & 0.56 $\pm$ 0.01 \\ 
\hline
&&&& Our Re-analysis \\
\hline

SBS 0335-052 & 15,940 $\pm$ 2,710 & 347 $^{+942}_{-347}$ & 4.6 $\pm$ 0.9 & 0.1 
$^{+0.2}_{-0.1}$ & 
0.0763 $\pm$ 0.0049 & 0.0023 $\pm$ 0.0023 & 0.2391 $\pm$ 0.0116 & 0.19 $\pm$ 0.01 \\ 

NGC 2363A    & 14,040 $\pm$ 1,060 & 285 $^{+343}_{-285}$ & 1.4 $\pm$ 0.4 & 
0.35 $\pm$ 0.28 & 
0.0801 $\pm$ 0.0046 & 0.0008 $\pm$ 0.0008 & 0.2441 $\pm$ 0.0107 & 0.69 $\pm$ 0.01 \\

SBS 0940+544N& 19,260 $\pm$ 2,480 & 35  $^{+140}_{-35}$ & 0.4 $\pm$ 0.4 & 
0.1 $^{+0.3}_{-0.1}$ & 
0.0841 $\pm$ 0.0035 & 0.000 & 0.2516 $\pm$ 0.0078 & 0.26 $\pm$ 0.01 \\ 

MRK 193      & 12,790 $\pm$ 1,760 & 820 $^{+860}_{-820}$ & 
1.9 $\pm$ 0.7  & 0.0 $^{+0.1}_{-0.0}$ & 
0.0802 $\pm$ 0.0046 & 0.0011 $\pm$ 0.0011 & 0.2451 $\pm$ 0.0107 & 0.61 $\pm$ 0.02 \\ 

SBS 1159+545 & 19,330 $\pm$ 2,100 &  75 $^{+116}_{-75}$ & 
0.5 $\pm$ 0.4  & 0.0 $^{+0.1}_{-0.0}$ &
0.0838 $\pm$ 0.0031 & 0.0010 $\pm$ 0.0010 & 0.2531 $\pm$ 0.0073 & 0.30 $\pm$ 0.01 \\ 

HARO 29      & 16,680 $\pm$ 890 &  33 $^{+97}_{-33}$ & 0.4 $\pm$ 0.2 & 
0.5 $\pm$ 0.2  & 
0.0844 $\pm$ 0.0024 & 0.0010 $\pm$ 0.0010 & 0.2543 $\pm$ 0.0058 & 0.58 $\pm$ 0.01 \\ 


SBS 1420+544 & 20,370 $\pm$ 1,740 &  27 $^{+83}_{-27}$ & 3.3 $\pm$ 0.4 & 
0.1 $^{+0.2}_{-0.1}$  & 
0.0854 $\pm$ 0.0026 & 0.0011 $\pm$ 0.0011 & 0.2568 $\pm$ 0.0062 & 0.54 $\pm$ 0.01 \\ 
\enddata
\end{deluxetable}

\newpage

\begin{deluxetable}{lcccccccc}
\tabletypesize{\footnotesize}
\rotate
\tablenum{12}
\tablewidth{0pt}
\tablecaption{Summary of Re-Analysis of IT04 ``High Quality'' Data Sample \label{tbl-12}}
\tablehead{
\colhead{Name} &
\colhead{T (K)} &
\colhead{n (cm$^{-3}$)} &
\colhead{$\tau$(3889)} &
\colhead{He(ABS) (EW)} &
\colhead{He$^+$/H$^+$} &
\colhead{He$^{++}$/H$^+$} &
\colhead{Y} &
\colhead{O/H $\times$ 10$^5$}
}
\startdata
&&&& IT04 \\
\hline
J 0519+0007  & 20,740 $\pm$ 340 &    ...       &    ...        & 0 &
0.0783 $\pm$ 0.0015 & 0.0023 $\pm$ 0.0002 & 0.2437 $\pm$ 0.0047 & 2.70 $\pm$ 0.09 \\ 

HS 2236+1344 & 21,120 $\pm$ 290 &    ...       &     ...       & 0 &
0.0789 $\pm$ 0.0013 & 0.0010 $\pm$ 0.0001 & 0.2421 $\pm$ 0.0040 & 2.96 $\pm$ 0.08 \\

HS 0122+0743 & 17,740 $\pm$ 230 &    ...       &     ...       & 0 &
0.0826 $\pm$ 0.0013 & 0.0007 $\pm$ 0.0001      & 0.2497 $\pm$ 0.0042 & 3.97 $\pm$ 0.11 \\ 

HS 0837+4717 & 19,510 $\pm$ 240 &    ...       &     ...       & 0 &
0.0763 $\pm$ 0.0011 & 0.0021 $\pm$ 0.0001 & 0.2386 $\pm$ 0.0034 & 3.98 $\pm$ 0.10 \\ 

CGCG 007-025 & 16,470 $\pm$ 170 &    ...       &     ...       & 0 &
0.0786 $\pm$ 0.0009 & 0.0012 $\pm$ 0.0001 & 0.2417 $\pm$ 0.0029 & 5.96 $\pm$ 0.14 \\ 

HS 0134+3415 & 16,390 $\pm$ 180 &     ...      &     ...       & 0 &
0.0812 $\pm$ 0.0013 & 0.0005 $\pm$ 0.0002 & 0.2459 $\pm$ 0.0042 & 7.20 $\pm$ 0.19 \\
 
HS 1028+3843 & 15,820 $\pm$ 160 &    ...       &     ...       & 0 &
0.0800 $\pm$ 0.0011 & 0.0012 $\pm$ 0.0001 & 0.2449 $\pm$ 0.0035 & 7.81 $\pm$ 0.20 \\ 

\hline
&&&& Our Re-analysis \\
\hline

J 0519+0007  & 22,050 $\pm$ 1,910 & 335 $\pm$ 189 & 2.09 $\pm$ 0.86 & 
0.18 $\pm$ 0.27 &
0.0799 $\pm$ 0.0070 & 0.0022 $\pm$ 0.0022 & 0.2471 $\pm$ 0.0166 & 2.46 $\pm$ 0.10 \\ 

HS 2236+1344 & 22,740 $\pm$ 2,230 & 139 $\pm$ 226 & 4.52 $\pm$ 0.71 & 
0.38 $\pm$ 0.40 &
0.0852 $\pm$ 0.0057 & 0.0009 $\pm$ 0.0009 & 0.2560 $\pm$ 0.0129 & 2.76 $\pm$ 0.09 \\

HS 0122+0743 & 18,860 $\pm$ 2,210 & 31 $\pm$ 107  & 0.72 $\pm$ 0.43 & 
1.19 $\pm$ 0.32 &
0.0887 $\pm$ 0.0041 & 0.0007 $\pm$ 0.0007 & 0.2632 $\pm$ 0.0089 & 3.81 $\pm$ 0.12 \\       

HS 0837+4717 & 21,850 $\pm$ 1,760 & 303 $\pm$ 112 & 4.36 $\pm$ 0.61 & 
0.01 $\pm$ 0.03 &
0.0835 $\pm$ 0.0038 & 0.0020 $\pm$ 0.0020 & 0.2547 $\pm$ 0.0095 & 3.64 $\pm$ 0.13 \\ 

CGCG 007-025 & 20,180 $\pm$ 2,210 & 85 $\pm$ 116 & 0.87 $\pm$ 0.49 & 
0.43 $\pm$ 0.30 &
0.0894 $\pm$ 0.0051 & 0.0012 $\pm$ 0.0012 & 0.2657 $\pm$ 0.0114 & 5.69 $\pm$ 0.19 \\ 

HS 0134+3415 & 18,160 $\pm$ 2.370 & 108 $\pm$ 205 & 1.06 $\pm$ 0.58 & 
0.25 $\pm$ 0.31 &
0.0840 $\pm$ 0.0051 & 0.0005 $\pm$ 0.0005 & 0.2523 $\pm$ 0.0115 & 6.86 $\pm$ 0.20 \\ 

HS 1028+3843 & 17,530 $\pm$ 2,940 & 461 $\pm$ 448 & 5.35 $\pm$ 0.88 & 
0.03 $\pm$ 0.14 &
0.0880 $\pm$ 0.0050 & 0.0012 $\pm$ 0.0012 & 0.2626 $\pm$ 0.0113 & 7.55 $\pm$ 0.23 \\ 

\enddata
\end{deluxetable}

\clearpage

\end{document}